 \documentclass[final,5p,times,twocolumn,authoryear]{elsarticle}

\usepackage{amssymb}
\usepackage{lipsum}
\usepackage{amsthm,amsmath}

\usepackage[breaklinks=true,colorlinks=true,
            linkcolor=blue,
            citecolor=blue,
            urlcolor=blue]{hyperref}

\usepackage{listings}
\usepackage{color,ulem}
\DeclareRobustCommand{\erase}{\bgroup\markoverwith{\textcolor{red}{\rule[.5ex]{2pt}{0.4pt}}}\ULon}

\journal{Astronomy $\&$ Computing}

\begin{document}

\begin{frontmatter}



\title{CROMA: a GPU-accelerated Fokker–Planck solver for modeling non-thermal emission in post-processing cosmological simulations}



\author[IRA]{Kosuke Nishiwaki}
\author[IRA]{Claudio Gheller}
\author[IRA]{Gianfranco Brunetti}
\author[ICRR]{Katsuaki Asano}

\affiliation[IRA]{organization={Istituto di Radioastronomia, INAF, Via Gobetti 101, 40121 Bologna, Italy}}


\affiliation[ICRR]{organization={Institute for Cosmic-ray Research, The University of Tokyo, 5-1-5, Kashiwanoha, Kashiwa, Chiba, 277-8582, Japan}}

\begin{abstract}
Numerical modeling of cosmic-ray (CR) spectra and the resulting non-thermal emission is important for understanding the physical mechanisms of CR acceleration and transport in astrophysical environments.
Galaxy clusters and cosmic filaments provide particularly important environments for studying these processes, where shocks, turbulence, magnetic fields, and CRs jointly shape diffuse synchrotron emission. 
Spectrally resolved modeling of CRs, including cooling, re-acceleration, and hadronic interactions, requires solving time-dependent Fokker--Planck (FP) equations.
An increasing number of studies employ FP solvers coupled with magnetohydrodynamical (MHD) simulations to follow the evolution of CRs in dynamically evolving environments.

We present a new C/C++-based FP solver with GPU acceleration for post-processing particle-based outputs from cosmological MHD simulations. 
The code evolves CR proton and electron spectra along Lagrangian tracer trajectories and consistently includes re-acceleration and secondary-particle production through hadronic interactions.
With the Chang--Cooper finite-difference scheme, the FP equation for each tracer is reduced to an independent tridiagonal system, which is solved on NVIDIA GPUs using a CUDA implementation of the parallel cyclic reduction (PCR) algorithm. 
The code uses sub-stepping between tracer snapshots, with the timestep constrained by the relevant CR cooling or re-acceleration timescale, and a batching algorithm that groups tracers with similar sub-stepping patterns for efficient execution.
We find that the GPU acceleration increases the per-node throughput by a factor of $\approx$10 on the Leonardo Booster system operated by the Italian National Supercomputing Centre (CINECA).
The fast FP solver is useful for high-resolution simulations dedicated to comparison with next-generation radio observations, extensive surveys of uncertain sub-grid acceleration models, and future multi-messenger applications combining radio, gamma-ray, and high-energy neutrino observations.
\end{abstract}






\end{frontmatter}




\section{Introduction}
\label{introduction}

Combining magnetohydrodynamical (MHD) simulations with numerical modeling of particle acceleration and transport is essential for studying the origin and evolution of cosmic rays (CRs) and their interplay with the thermal plasma.
Galaxy clusters provide a particularly important environment for studying these processes \citep[see][for reviews]{Dolag_2008,Kravtsov_Borgani_2012,Brunetti_Jones_review}.
Radio observations have revealed diffuse synchrotron emission extending over megaparsec (Mpc) scales, indicating the presence of relativistic electrons and magnetic fields in the intra-cluster medium (ICM) \citep[e.g.,][]{vanWeeren_review}.
Hadronic interactions of CR protons can produce secondary particles as well as gamma-ray and high-energy neutrino emission, which can be tested by current and future observations such as the Cherenkov Telescope Array (CTA) and IceCube \citep[e.g.,][]{Fang_Murase_2018,Nishiwaki_2021,Hussain_2023}.
Those observations provide a probe of CR acceleration processes, magnetic-field amplification, and their connection to the dynamical evolution of the large-scale structure of the Universe.
\par

CRs in the ICM can be accelerated through first- and second-order Fermi mechanisms, associated respectively with shocks and turbulence driven by mergers of clusters.
However, turbulence and shocks in the ICM are considered too weak to efficiently accelerate particles directly from the thermal pool.
Hence, current models of diffuse synchrotron emission, such as radio halos and radio relics, often invoke the re-acceleration of pre-existing relativistic particles accumulated over the lifetime of clusters through activity associated with active galactic nuclei (AGNs), galaxies, and cosmological shocks \citep[e.g.,][]{Voelk_1996,Berezinsky_1997,Brunetti_2001,Peterson_2001,Fujita_2003}.
In those models, the observable radio emission depends not only on the instantaneous properties of shocks and turbulence, but also on the time-dependent spectral evolution of CRs under cooling, injection, re-acceleration, and secondary-particle production through hadronic interactions.
\par

Recent observations with next-generation radio interferometers, such as LOFAR and MeerKAT, have revealed increasingly complex morphologies and spectral structures in diffuse radio emission, including spatial variations in spectral indices and correlations between radio and thermal emission.
These observations highlight features that are difficult to capture with simplified models, motivating more self-consistent comparisons between observations and multi-dimensional numerical simulations \citep[e.g.,][]{Rajpurohit_2021,Botteon_2022_A2255,vanWeeren_2024,vanWeeren_2026_A1795}.
\par

An increasing number of studies have coupled CR spectral-evolution models with idealized or cosmological MHD simulations.
In particular, time-dependent Fokker--Planck (FP) solvers have been used to follow the evolution of CR spectra, either in post-processing or through on-the-fly implementations \citep[][]{Winner_2019,Vazza_2023,Boess_2023,Beduzzi_2024,2025ApJ...989..155K,Ivleva_2026,Dominguez-Fernandez_2026,Nishiwaki26}.
Such simulations are powerful tools for interpreting current radio observations and for making predictions for future surveys with the Square Kilometre Array (SKA) and its precursors and pathfinders.
\par

As the mean free path of CRs is much shorter than the spatial grid in MHD simulations, sub-grid models are required to calculate the re-acceleration and injection processes due to wave-particle interaction in the FP equation (Sect.~\ref{sec:FP}). 
This is implemented with a momentum diffusion coefficient and injection term as a function of the outputs of MHD simulations. 
As observational constraints improve and numerical studies of plasma turbulence advance, these sub-grid models are expected to become increasingly sophisticated.
However, such simulations remain computationally demanding.
\par

The physical timescales involved in the FP equation can be much shorter than the MHD timescale, requiring many iterations to evolve CR spectra accurately.
In addition, extensive parameter surveys are often needed to explore sub-grid models of CR acceleration and transport, further increasing the computational cost.

In this work, we address this computational challenge by presenting a new C/C++-based GPU-accelerated FP solver, which we name CROMA.
Our code uses tracer particle outputs from MHD simulations such as Enzo \citep[][]{Bryan_2014} and follows the spectral evolution of CRs along the tracer trajectories, where each tracer carries an independent CR spectrum neglecting spatial diffusion. 
Several FP solvers have been developed for similar applications, including CREST \citep[][]{Winner_2019}, ROGER \citep[][]{Vazza_2023}, and CRESCENDO \citep[][]{Boess_2023}. 
On the physical side, our code consistently models spectra of both CR protons and electrons with CR re-acceleration and secondary-particle production through hadronic interactions.
On the computational side, the code has a CUDA-based GPU backend, enabling parallel computation of a large number of independent FP problems.
These features make the code particularly well suited for future applications in multi-messenger astronomy and for high-resolution simulations designed for comparison with next-generation radio observations.
\par

The CUDA backend uses the parallel cyclic reduction (PCR) algorithm to solve the tridiagonal systems obtained by discretizing the FP equation in momentum space (Sect.~\ref{sec:code}).
By assigning each tracer-level FP problem to a CUDA thread block and parallelizing the operations over momentum bins, the implementation makes effective use of the large number of threads available on GPUs (\ref{app:CC}).
We test the code on the Leonardo supercomputing system at CINECA and evaluate its performance and scalability for large numbers of tracer particles (Sect.~\ref{sec:scalability}).
We also discuss future applications of the code, including CR transport between tracer particles and run-time CR spectral evolution in cosmological simulations (Sect.~\ref{sec:conclusions}).
The code developed in this work is publicly available on GitHub at \url{https://github.com/kosukenishiwaki/croma-fp}.
\par

\section{Fokker--Planck equation}
\label{sec:FP}
We aim to model the spectral evolution of CRs under energy-loss and acceleration processes.
We consider one-dimensional spectra in momentum space for CR electrons $\mathcal{N}_e(p)$ and CR protons $\mathcal{N}_p(p)$ represented by a tracer particle, where $p$ is the CR particle momentum.
In what follows, we omit the species subscript of $\mathcal{N}$ unless it is explicitly required.
Here, $\mathcal{N}(p)\,dp$ represents the number of CR particles associated with the tracer in the momentum interval $[p,p+dp]$. 
The total number of CR particles carried by a tracer is calculated as $\int_{p_{\rm min}}^{p_{\rm max}}\mathcal{N}(p)dp$, where $p_{\rm min}$ and $p_{\rm max}$ are minimum and maximum momenta, respectively.
The corresponding per-volume spectrum can be obtained by dividing
$\mathcal{N}(p)$ by the effective volume of each tracer.
Neglecting spatial diffusion, the time evolution of $\mathcal{N}(p)$ is modeled with the FP equation of the following form:
\begin{align}
    \frac{\partial \mathcal{N}(p)}{\partial t} &= \frac{\partial}{\partial p}\left[D_{pp}\frac{\partial \mathcal{N}(p)}{\partial p}+\left(H(p,z(t))-\frac{2}{p}D_{pp}\right)\mathcal{N}(p) \right] \nonumber\\
    & \; - \frac{\mathcal{N}}{T(p)} + Q(p,z(t)),
    \label{eq:FP}
\end{align}
where $H$, $D_{pp}$, $T$, and $Q$ denote cooling, momentum diffusion, sink, and injection coefficients, respectively. 
For CR electrons, the cooling term includes radiative losses due to synchrotron and inverse-Compton emission, Coulomb cooling, and adiabatic compression or expansion, while Coulomb cooling and adiabatic compression or expansion are considered for CR protons
\citep[see][for details]{Nishiwaki_2021,Nishiwaki26}.
\par

For stochastic reacceleration represented with $D_{pp}$, the code supports the sub-grid models conventionally used for diffuse radio emission in clusters; Alfv\'enic stochastic acceleration \citep[ASA,][]{BL16} and transit-time damping \citep[TTD,][]{BL07,Brunetti_2016}.
In those models, the reacceleration efficiency is expressed as a function of the turbulent Mach number, $\mathcal{M}_{\rm turb} = \delta v/c_{\rm s}$, where $c_{\rm s}$ is the sound speed.

The sink term is neglected for CR electrons. For CR protons, the sink term represents the loss through the hadronic $pp$ collision with the background baryonic matter (Sect.~\ref{sec:hadronic}).

The source term $Q$ should be determined from the physical assumption for the CR injection. In the default setup, the code sets a power-law spectrum at the beginning of the simulation and neglects additional sources by setting $Q = 0$. 
When CR protons are included, the code follows the continuous injection of secondary electrons through the hadronic interaction (Sect.~\ref{sec:hadronic}).
In future work, we plan to extend the source term to include the effect of diffusive shock acceleration (DSA) at shocks, following previous studies \citep[e.g.,][]{Boess_2023,Vazza_2023}.

\subsection{Chang-Cooper finite difference scheme}
\label{sec:Chang-Cooper}

We set a one-dimensional momentum grid with $N_{\rm bin}$ bins equally-spaced in logarithmic scale. 
The number of momentum bins must be a power of two, $N_{\rm bin} =  2^M$, so that the PCR solver is applicable in the CUDA backend (see also \ref{app:PCR}).
We adopt the Chang--Cooper (CC) discretization for the momentum-space flux \citep{Chang_Cooper_70,Park_Petrosian_96} and use a first-order implicit time integration, 
\begin{equation}
    \frac{\mathcal{N}_j^{i+1} - \mathcal{N}_j^{i}}{\Delta t} = \frac{F_{j+\frac{1}{2}}^{i+1} - F_{j-\frac{1}{2}}^{i+1}}{\Delta p_j} -\frac{\mathcal{N}_j^{i+1}}{T_j} + Q_j,
    \label{eq:FP_CC}
\end{equation}
where $\Delta t$ is the discretized time step and we use index $i$ and $j$ for the time step and the momentum bin, respectively.
The continuous expression of the flux is
\begin{equation}
    F(p,t) = D_{pp}(p,t)\frac{\partial \mathcal{N}(p,t)}{\partial p} - \left(H(p,t) - \frac{2D_{pp}(p)}{p}\right)\mathcal{N}(p,t).
    \label{eq:CC_flux}
\end{equation}


In short, the CC prescription evaluates the fluxes, $F_{j+\frac{1}{2}}$ and $F_{j-\frac{1}{2}}$, at the interface between adjacent momentum bins by interpolating the value of $\mathcal{N}_j^i$ from the two neighboring bin centers, with the weight $\delta_{j}$ for advection determined by the local ratio of advective drift to diffusion.
The weight $\delta_{j}$ is chosen to preserve the correct local balance between drift and diffusion in the steady state.
It reduces to a centered difference when diffusion dominates and approaches an upwind discretization when the drift term dominates (see \citet{Park_Petrosian_96} for details).


With this scheme, Eq.~(\ref{eq:FP_CC}) leads to a form of a tridiagonal linear system,
\begin{equation}
    -a_j \mathcal{N}_{j-1}^{i+1} + b_j \mathcal{N}_j^{i+1} - c_j \mathcal{N}_{j+1}^{i+1}
    = \Delta t\,Q_j + \mathcal{N}_j^i.
\label{eq:tridiag}
\end{equation}
Here, we omit the specific expressions for coefficients $a_j$, $b_j$, and $c_j$. We adopt the no-flux boundary conditions
\begin{equation}
a_0=0, \;\;\;\;  c_{N_{\rm bin}-1}=0. 
\end{equation}
With the no-flux boundary conditions, the CC discretization conserves particle number in the absence of source and sink terms and preserves the positivity of the solution.
\par

\subsection{Hadronic chain and secondary particles}
\label{sec:hadronic}
The code calculates the secondary particle creation through the decay of pions generated through the hadronic $pp$ interaction of CR protons:
\begin{eqnarray}
\left\{
\begin{array}{l}
p+p\rightarrow \pi^{0,\pm} + X, \\
\pi^0 \rightarrow 2\gamma,\;\;\pi^\pm \rightarrow \mu^\pm +\nu_\mu (\bar{\nu}_\mu),   \\
\mu^\pm \rightarrow e^\pm+\bar{\nu}_\mu(\nu_\mu) +\nu_\mathrm{e} (\bar{\nu}_\mathrm{e}).
\end{array}
\right.
\label{eq:hadronic}
\end{eqnarray}
The spectra of the pions and secondary particles are calculated with the approximate expression given in \citet{Kelner_2006}.
The secondary electrons and positrons are added to the CR electron spectrum through the source term $Q$ in Eq.~(\ref{eq:FP}).
The code uses precomputed tables for the secondary spectra as a function of the momenta of the parent CR proton and secondary particles \citep[see][for further details]{Nishiwaki_2021}.
\par

The energy loss of CR protons through $pp$ collisions is approximated with the sink term as 
\begin{equation}
    T(p) = \left[ n_{\rm gas}\sigma_{pp}(p)c \right]^{-1},
\end{equation}
where $n_{\rm gas}$ is the number density of the thermal gas particles, and $\sigma_{pp}$ is the total inelastic cross section of the $pp$ collision \citep[][]{Blattnig_2000,Kamae_2006}. 
The code optionally calculates the emissivity of the secondary gamma ray and neutrinos for each tracer at each snapshot.

\section{Synchrotron emission with precomputed tables}
\label{sec:synch}

The code calculates the synchrotron emissivity of CR electrons in each tracer particle. 
Under the assumption of isotropic pitch-angle distribution for CR electrons, the synchrotron emissivity per unit frequency per unit solid angle can be written as
\begin{equation}
j_\nu (\nu,B;\mathcal{N}_e)=
\frac{\sqrt{3}\,e^3 B}{8\pi m_e c^2}
\int_{p_{\rm min}}^{p_{\rm max}} dp \, \mathcal{N}_e(p)
\int_0^\pi d\theta \, \sin^2\theta \,
F\!\left(\frac{\nu}{\nu_{\rm c}}\right),
\label{eq:synch}
\end{equation}
where
\begin{equation}
F(x)=x\int_x^\infty K_{5/3}(\xi)\,d\xi,
\label{eq:K}
\end{equation}
is the synchrotron kernel and
\begin{equation}
\nu_{\rm c}(p,B,\theta)=
\frac{3 e B}{4\pi m_e c}\gamma^2 \sin\theta ,
\label{eq:nu_c}
\end{equation}
with $\gamma=\sqrt{1+(p/m_ec)^2}$.
\par

The calculation of Eq.~(\ref{eq:synch}) involves three numerical integrals and can be computationally demanding.
We circumvent this problem by using pre-computed numerical tables.
To accelerate the calculation of Eqs.~(\ref{eq:K}), we prepare a table for the synchrotron kernel $F(x)$ on a uniform grid in $\log x$. 
We use $N_x = 2750$ logarithmic bins between $\log_{10} x_{\rm min} = -27.5$ and $\log_{10} x_{\rm max} = 2.5$.
\par

The integration over the pitch angle $\theta$ in Eq.~(\ref{eq:synch}) can also be accelerated using fixed binning for $\theta$ and the frequency $\nu$ as $\theta_k$ and $\nu_l$, respectively. 
To this end, we tabulate the characteristic frequency $\nu_{\rm c}$ at unit magnetic field,
\begin{equation}
\nu_{{\rm c},jk}^{(0)} \equiv \nu_{\rm c}(p_j, B_0, \theta_k),
\end{equation}
and define
\begin{equation}
y_{jkl} \equiv
\log_{10}\!\left(
\frac{\nu_l}{\nu_{{\rm c},jk}^{(0)}}
\right),
\end{equation}
where $B_0=10^{-6}\,{\rm G}$ is the reference field. 
The binning for $\nu$ and $\theta$ can be tuned through parameters. 
By default, we use $N_\nu = 128$ bins for $10^{5.5}~{\rm Hz}<\nu<10^{10.5}~{\rm Hz}$ and $N_\theta = 512$ bins for $0<\theta<\pi$.
For a local magnetic field $B$, the parameter $x=\nu/\nu_{\rm c}$ for the synchrotron kernel $K(x)$ is obtained from
\begin{equation}
\log_{10} x_{jkl} (B) = y_{jkl}-\log_{10}\!\left(\frac{B}{B_0}\right).
\end{equation}
Thus, the magnetic-field dependence can be taken into account as a shift in $\log x$, while the $(p_j,\theta_k,\nu_l)$ dependence is stored in the table of $y_{jkl}$.
\par

The integral over $\theta$ in Eq.~(\ref{eq:synch}) is now a function of the magnetic field $B$.
One can assume a typical range of $B$ in the region of interest in astrophysical environments.
Setting a grid of $\log_{10} B$ in advance, we compute a table of the pitch-angle integrated kernel,
\begin{equation}
\tilde{F}_{bjl} \equiv
\sum_{k=0}^{N_\theta-1}
w_k\,
F\!\left(
x_{jkl}(B_b)
\right),
\label{eq:synch_table}
\end{equation}
where $w_k = \sin^2\theta_k \Delta\theta_k$ is the weight factor and $B_b$ denotes the magnetic field in the $b$-th $\log B$ bin.
We use the range $-6.0 \leq \log_{10}(B_{\mu{\rm G}}) \leq 2.0$, which sufficiently covers the magnetic-field strengths expected in galaxy clusters and cosmic filaments.

Note that the above $B$ table is used only for the synchrotron emission, and that the radiative cooling of CR electrons in the FP equation (Eq.~(\ref{eq:FP})) is calculated with the exact value of the local $B$ recorded in the tracer data.
Finally, the synchrotron emissivity is evaluated as
\begin{equation}
j_{\nu}(\nu_l,B;\mathcal{N}_e) =
\frac{\sqrt{3}\,e^3 B}{8\pi m_e c^2}
\sum_j \mathcal{N}_{e,j}\,\Delta p_j\,\tilde{F}_{bjl},
\end{equation}
with linear interpolation of $\log_{10}\tilde{F}$ between adjacent $\log B$ bins.

In \ref{app:synch_table}, we assess the errors involved in the tabulated method explained in this Section. 
Compared to the direct numerical integration, the error is smaller than $\mathcal{O}(10^{-3})$ over a wide frequency range, except at the far end of the exponential tail.
\par

\section{Code overview}
\label{sec:code}

\begin{figure}
    \centering
    \includegraphics[width=\linewidth]{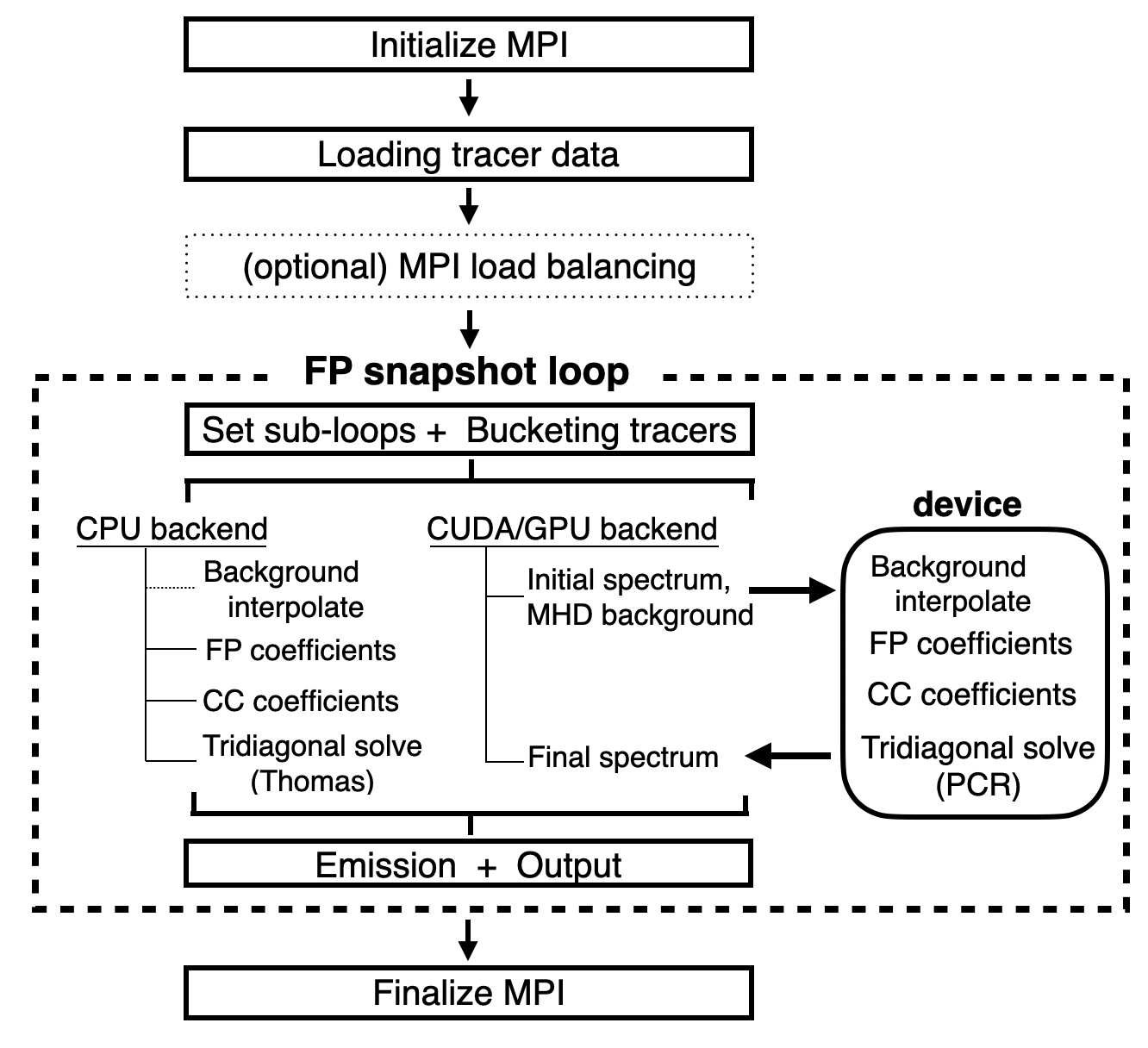}
    \caption{Schematic picture of the code workflow. The CUDA backend path involves the communication to the GPU device.}
    \label{fig:flow}
\end{figure}

Figure~\ref{fig:flow} shows a schematic overview of the code workflow.
The code first reads tracer snapshot data and distributes the tracer workload across the CPU cores with the Message Passing Interface (MPI) \citep[][]{gropp2014}.
Between adjacent snapshots, the FP equation is solved with internal sub-stepping, where the timestep is constrained by the shortest relevant CR cooling or reacceleration timescale (Sect.~\ref{sec:timestep}).

The background MHD quantities are linearly interpolated between adjacent snapshots during the sub-stepping. 
When the number of sub-steps becomes very large, the interpolation at every sub-step can be unnecessarily expensive. 
In such a case, the code allows the interpolation to be evaluated on a coarser temporal grid, with the maximum number of interpolation updates between two snapshots controlled by a runtime parameter.

\par

An optional MPI load-balancing step redistributes tracers using the estimated number of FP sub-steps as a proxy for the computational cost of each tracer (Sect.~\ref{sec:loadbalance}).
Within each rank, tracers with the same number of sub-steps are grouped together and advanced using batched FP solves (Sect.~\ref{sec:batched}).
The code has two backends for this batched process: a CPU backend and a CUDA backend.
The CPU backend solves the resulting batched tridiagonal systems with the classical Thomas algorithm because of its small work complexity.
The CUDA backend uses PCR, because it can update the momentum bins of many tracers in parallel and efficiently utilize the large number of GPU threads, allowing faster execution for large batches of independent systems (\ref{app:PCR}).

The code outputs the CR spectra and associated synchrotron emission spectra for each tracer at each snapshot.
When CR protons are included in the calculation, the code can also output hadronic gamma ray and neutrino spectra.

\subsection{Input data}
\label{sec:input}
The code is designed to post-process tracer-particle data extracted from cosmological MHD simulations.
The input data are provided as Hierarchical Data Format version 5 (HDF5) files, with one file corresponding to each simulation snapshot.
Each file contains the local MHD quantities, such as the gas density, temperature, and magnetic field, sampled at the position of each tracer at each snapshot.
Each tracer particle should be identified by a unique ID that is preserved across snapshots, allowing its trajectory to be reconstructed in time within the code.
\par

When turbulent reacceleration models such as ASA or TTD (Sect.~\ref{sec:FP}) are adopted, the local turbulent velocity amplitude is required to calculate the reacceleration timescale.
The input data structure can be easily extended for future applications, for example to include shock-related quantities for modeling the diffusive shock acceleration.

The magnetic field strength in the original cosmological simulations may be underestimated when turbulent dynamo amplification is not fully resolved at the available spatial resolution.
To account for such cases, the code provides an option to replace the simulated magnetic field strength with an estimate based on the turbulent dynamo model adopted in \citet{Nishiwaki26}.
\par

The physical model and its parameters are controlled at runtime through a parameter file. The parameter file is supplied as a command-line argument when the code is executed.

\subsection{Time stepping}
\label{sec:timestep}

The computational cost of the post-processing calculation is primarily determined by the number of internal time steps used to solve the FP equation.
Although the tracer data are stored only at discrete simulation snapshots, the cooling and reacceleration timescales entering the FP model can be significantly shorter than the snapshot interval.
The code therefore introduces sub-stepping between adjacent tracer snapshots to accurately follow the momentum-space evolution of cosmic-ray particles.
\par

For each tracer and each snapshot interval, the target FP time step for sub-stepping, $\Delta t_{\rm FP}$, should satisfy the condition of
\begin{equation}
\Delta t_{\rm FP} \leq C_{\rm safe}
\min \left( t_{\rm cool}^{\rm min}, t_{\rm reacc} \right),
\label{eq:dt_fp}
\end{equation}
where $C_{\rm safe}<1$ is a safety factor, $t_{\rm cool}^{\rm min}$ is the minimum cooling timescale over the momentum grid, and $t_{\rm reacc} \equiv p^2/(4 D_{pp})$ is the local reacceleration timescale.
In the conventional hard-sphere-type reacceleration, as for ASA and TTD, $t_{\rm reacc}$ does not depend on the particle momentum, i.e., $D_{pp}\propto p^2$.
For CR electrons in the ICM, $t_{\rm cool}^{\rm min}$ is typically determined either by the Coulomb cooling at the minimum momentum $p_{\rm min}$ or by the radiative cooling at the maximum momentum $p_{\rm max}$.
Thus, the computational load depends on the choice of these values.
The cooling timescale of CR protons is typically much longer than that of CR electrons over most of the momentum range considered here, and therefore usually does not set the limiting time step.
\par

As explained in Sect.~\ref{sec:batched}, we adopt a batched method to evolve the FP timestep by grouping tracers with the same number of sub-steps.
However, assigning a completely independent value of $\Delta t_{\rm FP}$ to every tracer would fragment the tracer population into many small groups and reduce the efficiency of the batched method.
To avoid this fragmentation of tracer groups, the target FP time step $\Delta t_{\rm FP}$ is rounded down to a logarithmic mantissa ladder.
In the present implementation, this ladder has ten allowed mantissae per decade, ${1.0, 1.25, 1.6, 2.0, 2.5, 3.2, 4.0, 5.0, 6.4, 8.0}$.
Note that the rounding of $\Delta t_{\rm FP}$ to a smaller value does not violate the safety condition of Eq.~(\ref{eq:dt_fp}).

With the rounded $\Delta t_{\rm FP}$, we estimate the number of FP sub-steps between two adjacent tracer snapshots as
\begin{equation}
N_{\rm sub} =
\left\lceil \frac{\Delta t_{\rm snap}}{\Delta t_{\rm FP}} \right\rceil,
\label{eq:N_sub}
\end{equation}
where $\Delta t_{\rm snap}$ is the time interval between the two snapshots.
For each $N_{\rm sub}$ assigned to tracers, the actual FP timestep is determined as $\Delta t_{\rm FP} = \Delta t_{\rm snap}/N_{\rm sub}$.
During the sub-step loop, the background MHD quantities from the tracer files are linearly interpolated in time between two snapshots.


\begin{figure}
    \centering
    \includegraphics[width=\linewidth]{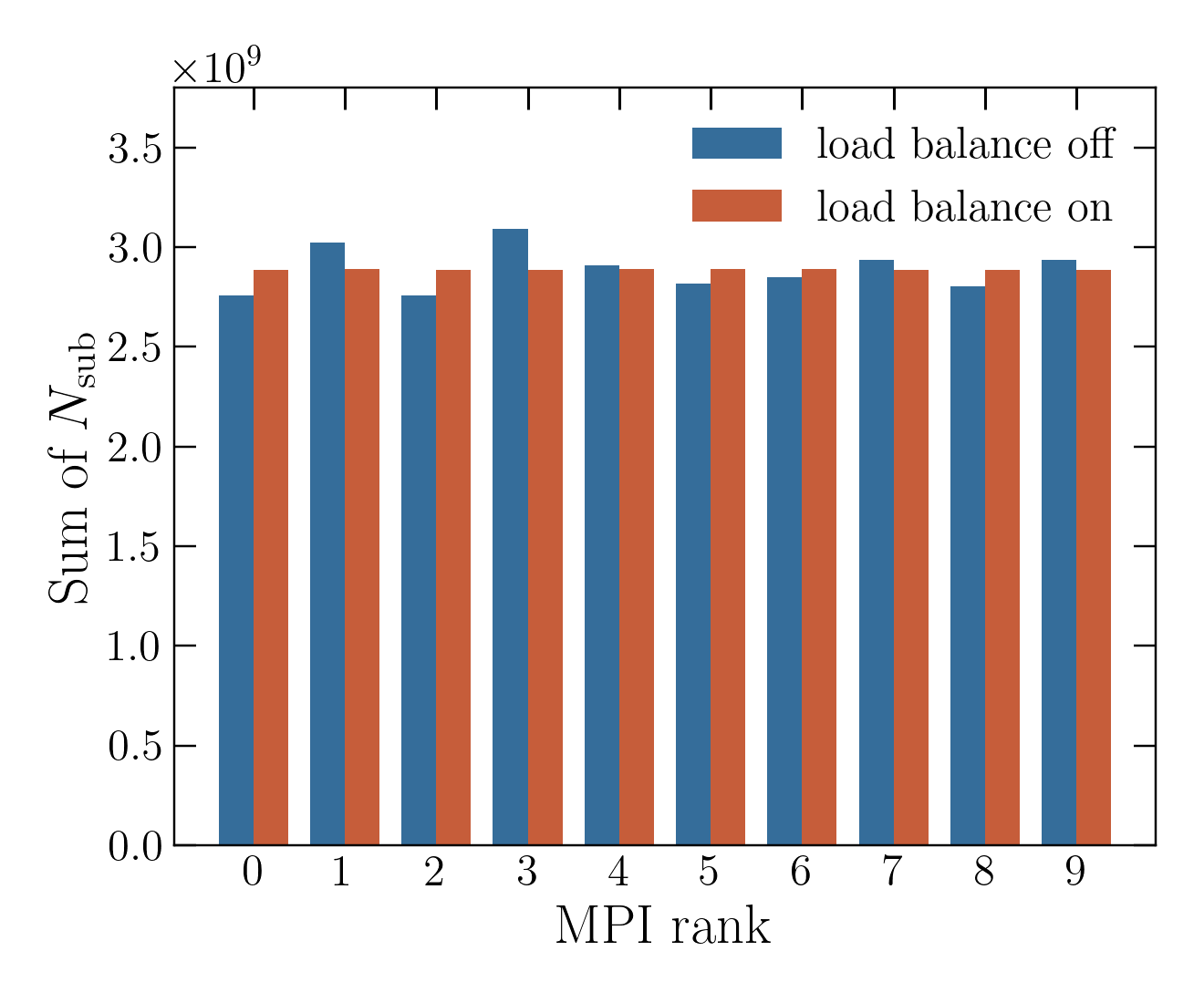}
    \caption{Distributions of the total number of sub-loops among the MPI ranks. The blue bars show the case with an equal number of tracer particles on each rank, whereas the red bars show the case with load balancing with respect to the total $N_{\rm sub}$, or $W_n$ (Eq.~\ref{eq:Wi}). Shown is the case for 10$^6$ tracer particles and ten MPI ranks.}
    \label{fig:load_balance}
\end{figure}

\subsection{Load balancing}
\label{sec:loadbalance}

The code supports MPI parallelization by distributing the global tracer list across MPI ranks.
Because the FP problems for individual tracer particles are independent and require no communication between MPI tasks during the solve, the computational cost of each tracer can be estimated prior to the full numerical integration.
We use the accumulated number of FP sub-steps $N_{{\rm sub},n}$ for the $n$-th particle over all snapshot intervals as a proxy for the tracer workload,
\begin{equation}
W_n = \sum_{\rm snapshots} N_{{\rm sub},n}.
\label{eq:Wi}
\end{equation}
The computational load assigned to each MPI rank is approximated by the sum of $W_n$ over the tracers owned by that rank.
\par
In general cases, a simple uniform distribution of tracers over MPI ranks can lead to a non-uniform workload distribution on MPI tasks, as shown in Fig.~\ref{fig:load_balance}.
Since the overall execution time is determined by the slowest MPI rank, such load imbalance directly degrades the parallel performance.
To address this problem, the code has a load-balancing option based on the estimated tracer workload $W_n$.
Before the start of the main FP solve, each MPI rank estimates $W_n$ for the tracers initially assigned to it.
These estimates are collected and used to identify the tracers with large $W_n$.
A fraction of these computationally-expensive tracers is then reassigned from heavily loaded ranks to lightly loaded ranks, so that the total estimated workload becomes more even across MPI ranks.
The distribution of the workload after the load-balancing is shown with the red bars in Fig.~\ref{fig:load_balance}.

The redistribution changes only the tracer ownership of each rank.
Each rank constructs an active tracer list from the reassigned tracer IDs and their
source offsets.
These identifiers are used to access the original input data and write the updated spectra back to their global locations.
Hence, the load-balancing communication is limited to tracer IDs, estimated workloads, and assignment records, while the FP solves remain local to each MPI rank.
\par

The load-balancing procedure introduces an additional minor overhead due to the planning and the MPI communication.
In practical applications, however, this overhead is negligible compared with the cost of the main FP solve.
For example, in the case of Fig.~\ref{fig:load_balance}, the load-balancing overhead is approximately 0.3 second.
Before load balancing, the maximum estimated workload among MPI ranks is about 7\% larger than the mean.
This moderate imbalance may lead to differences in the wall-time of several tens of seconds in the GPU runs (see also Sect.~\ref{sec:scalability}).
Thus, the cost of the load-balancing step is much smaller than the potential wall-time penalty caused by the non-uniform workload.



\subsection{Batched computation and CUDA backend}
\label{sec:batched}

The update of CR spectrum by solving the FP equation is performed for batches of tracers.
Within each MPI rank, the tracers with the same sub-stepping pattern are grouped together.
We refer to these groups of tracers as \emph{buckets}.
Using the implicit CC finite-difference scheme, the FP solve leads to a tridiagonal system in momentum space (Sect.~\ref{sec:Chang-Cooper}).
Thus, the main numerical operation for each bucket is a batched tridiagonal solve, consisting of many independent systems with the same momentum-grid size.
\par



In the CPU backend, bucketed batching primarily serves to regularize the data layout and reduce per-tracer control overhead; it does not change the computational complexity of the tridiagonal solves. The systems in a bucket remain independent and are solved separately using a classical Thomas algorithm.
The CPU backend supports shared-memory parallelism with OpenMP within each MPI rank.
The tracers assigned to each MPI rank can be further distributed among OpenMP threads and processed independently, including the bucket construction and FP solves.
However, running a scaling test of Sect.~\ref{sec:cpu_strong} for a fixed total number of 32 CPU cores, we find that varying the decomposition between MPI ranks and OpenMP threads results in only minor differences in the overall performance.
Thus, we adopt one OpenMP thread per MPI rank as the default configuration and use this setup throughout the performance tests presented below.
Further optimization of the OpenMP implementation is left for future work, as well as the exploration of OpenMP offloading to provide GPU acceleration beyond the current CUDA-specific backend.
\par


On the other hand, in the CUDA backend, buffer arrays for CR spectra are allocated on the GPU device for all tracers owned by a MPI rank.
To minimize the host--device data communication, the CR spectra of each MPI task are copied from the host to these buffer arrays only at the beginning of each snapshot interval.
The tracer buckets are processed by selecting the corresponding tracer entries from these device arrays.
The preparation of the FP coefficients and tridiagonal matrices, as well as the tridiagonal solves, is performed on the GPU.
The CUDA backend is designed for MPI runs with one MPI task per GPU.
We note that allocating more CPU cores to the job than the number of MPI tasks used in the run can sometimes be beneficial for achieving the best GPU performance on a compute node.
\par

We adopt a PCR algorithm for the tridiagonal solve in the CUDA backend.
In the PCR kernel, each tridiagonal system is assigned to one CUDA thread block, and the $N_{\rm bin}$ threads within the block are mapped to the momentum bins of that system (\ref{app:PCR}).
In the present implementation, we limit the momentum-grid size to $N_{\rm bin}\leq512$.
This limit is imposed to keep the number of threads per block and the shared-memory usage within a practical range for efficient execution on NVIDIA A100.
\par

Although PCR has a higher work complexity than the Thomas algorithm, $\mathcal{O}(N_{\rm bin}\log_2 N_{\rm bin})$ instead of $\mathcal{O}(N_{\rm bin})$, its fine-grained parallelism is well suited to GPUs and can outperform serial tridiagonal solvers for large batches of independent systems (\ref{app:PCR}).
After all FP sub-loops are computed for all tracer buckets, the updated spectra are copied back to host memory.
\par

During the development of the CUDA backend, ChatGPT and Codex were used to assist with the CUDA implementation, code review, and debugging.
We note that the physical processes included in the Fokker--Planck equation and the Chang--Cooper numerical scheme had already been implemented in our previous work \citep[e.g.,][]{Nishiwaki_2021}.
The algorithmic choices, validation tests, and performance evaluations presented in this work were performed and verified by the authors.
\par

In Sect.~\ref{app:test} and \ref{app:CC}, we check the consistency between the CPU and CUDA backends by solving test problems.
We refer the reader to Fig.~\ref{fig:spectra} for examples of the output CR and emission spectra.



\section{Scalability tests}
\label{sec:scalability}

In this section, we evaluate the GPU acceleration implemented in our code using tracer data generated by an Enzo simulation of a massive galaxy cluster (Nishiwaki et al. in prep.). 
We vary the size of the tracer dataset and the number of MPI ranks, and perform full FP runs, including data loading and output writing. In the following tests, we use tracer data consisting of 100 snapshots between redshifts $z=1$ and $z=0$.
Note that the computational cost changes significantly from snapshot to snapshot because the cooling and reacceleration timescales depend on redshift and on the local MHD quantities.
We vary the number of tracers, instead of the number of snapshots, to control the problem size in the scalability tests.
In this way, all test datasets retain the same redshift range from $z=1$ to $z=0$, while the number of independent FP problems is varied by sampling tracers from the full dataset.

It is useful to define the throughput of the code as
\begin{equation}
R \equiv \frac{N_{\rm trc}}{T N_{\rm ranks}},
\end{equation}
where $N_{\rm trc}$ is the total number of tracers of the run, $T$ is the run time, and $N_{\rm ranks}$ is the number of MPI tasks or the number of GPUs. 
We vary the problem size $N_{\rm trc}$ in the following tests to assess the scalability of the code.
Datasets with different $N_{\rm trc}$ are generated by randomly sampling tracers from the full tracer dataset produced by the Enzo simulation, which contains $N_{\rm trc}^{\rm tot} = 1.5\times10^{7}$ tracers.
\par

We use a fixed physical model and parameter set throughout these tests. 
Specifically, 
We set initial populations of CR protons and electrons at the initial snapshot ($z=1$) with the power-law spectra $\mathcal{N}\propto p^{-2.2}$.
We adopt the ASA turbulent reacceleration model using the local vorticity recorded in the tracer data \citep[][]{BL16}, and $C_{\rm safe} = 0.5$ for the safety factor (Eq.~\ref{eq:dt_fp}).
The minimum and maximum momenta are set to $p_{\rm min}/mc = 10^{-0.5}$ and $p_{\rm max}/mc = 10^{6}$ for both CR electrons and protons.

The initial energy density ratio between CR proton and electron is set to be one. 
We use $N_{\rm bin} = 128$ bins for the logarithmic momentum grid for both CR electrons and protons.
With the CC discretization, this resolution provides sufficient accuracy for the spectral evolution for the astrophysical applications (\ref{app:CC}).
\par


All the tests are conducted with High Performance Computing resources on the Booster partition of the Leonardo supercomputer\footnote{\url{https://www.hpc.cineca.it/systems/hardware/leonardo/}} at the CINECA Italian Supercomputing centre.
The Leonardo Booster comprises 3456 compute nodes, each equipped with a 32-core Intel Xeon Platinum 8358 CPU, 512 GB of DDR4 memory, and four NVIDIA Ampere A100 GPUs with 64 GB of HBM2e memory each.
\par

\subsection{CPU strong scaling}
\label{sec:cpu_strong}
\begin{figure}
    \centering
        \includegraphics[width=\linewidth]{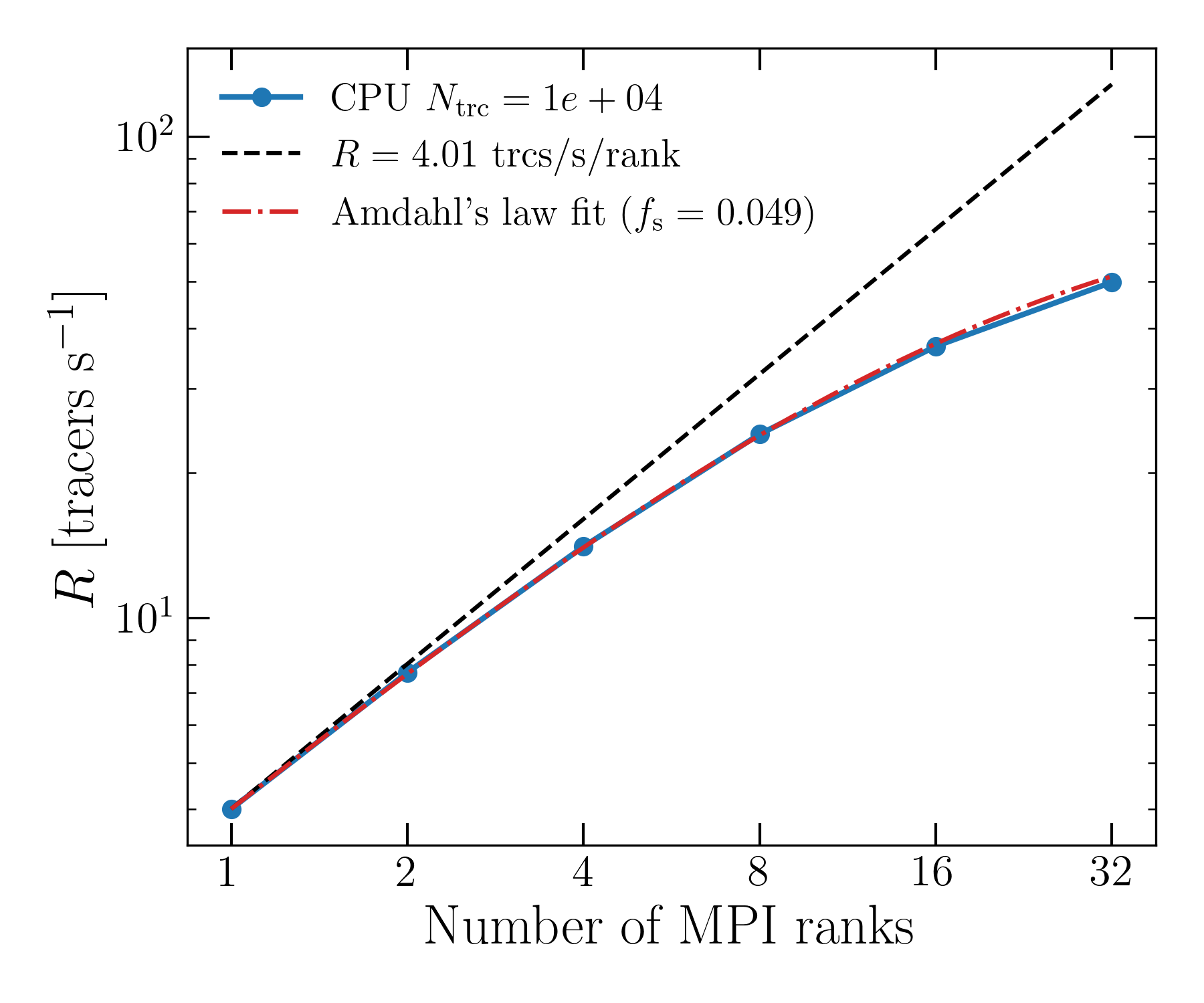}
    \caption{Strong scaling test with the CPU backend. The vertical axis shows the total throughput over all MPI ranks. Each MPI rank uses a single CPU thread in this benchmark. The test uses a fixed tracer sample with $N_{\rm trc} = 10^4$. The black dashed line shows the ideal scaling expected for a constant per-rank throughput.}
    \label{fig:cpu_strong}
\end{figure}

We first measure the performance of the FP solver with the CPU backend through a strong-scaling test.
We use a tracer sample with $N_{\rm trc} = 10^4$ tracers. 
This problem size gives a single-rank runtime of around forty minutes, which is long enough to distinguish the scalable part of the computation from fixed costs that are typically of order a few minutes.
We find that the end-to-end throughput is approximately $R\approx 4$ tracers s$^{-1}$ on a single rank.
We measure the speedup by increasing the number of MPI tasks from 1 to 32.
With 32 MPI tasks, each rank computes the FP equations for $\approx310$ tracers.
The run with 32 MPI tasks finishes in about three minutes, at which point the overhead associated with non-parallelized parts of the code significantly affects the total execution time.
\par

We fit the result with Amdahl's law and find an effective serial fraction of $f_{\rm s}= 0.049$.
This suggests that the maximum speedup available with this setup is approximately $\approx 20$.
In a realistic production run, the problem size can be as large as $N_{\rm trc}\sim10^7$ \citep[e.g.,][]{Nishiwaki26}.
The scaling test presented here shows that the speedup becomes limited when the workload falls below $\sim10^3$ tracers per MPI task.
Thus, for such production runs, using a few hundred MPI tasks would provide a reasonable balance between the total wall-clock time and the total CPU time consumption.


\subsection{GPU acceleration performance}
\label{sec:gpu_scaling}

\begin{figure}
    \centering
        \includegraphics[width=\linewidth]{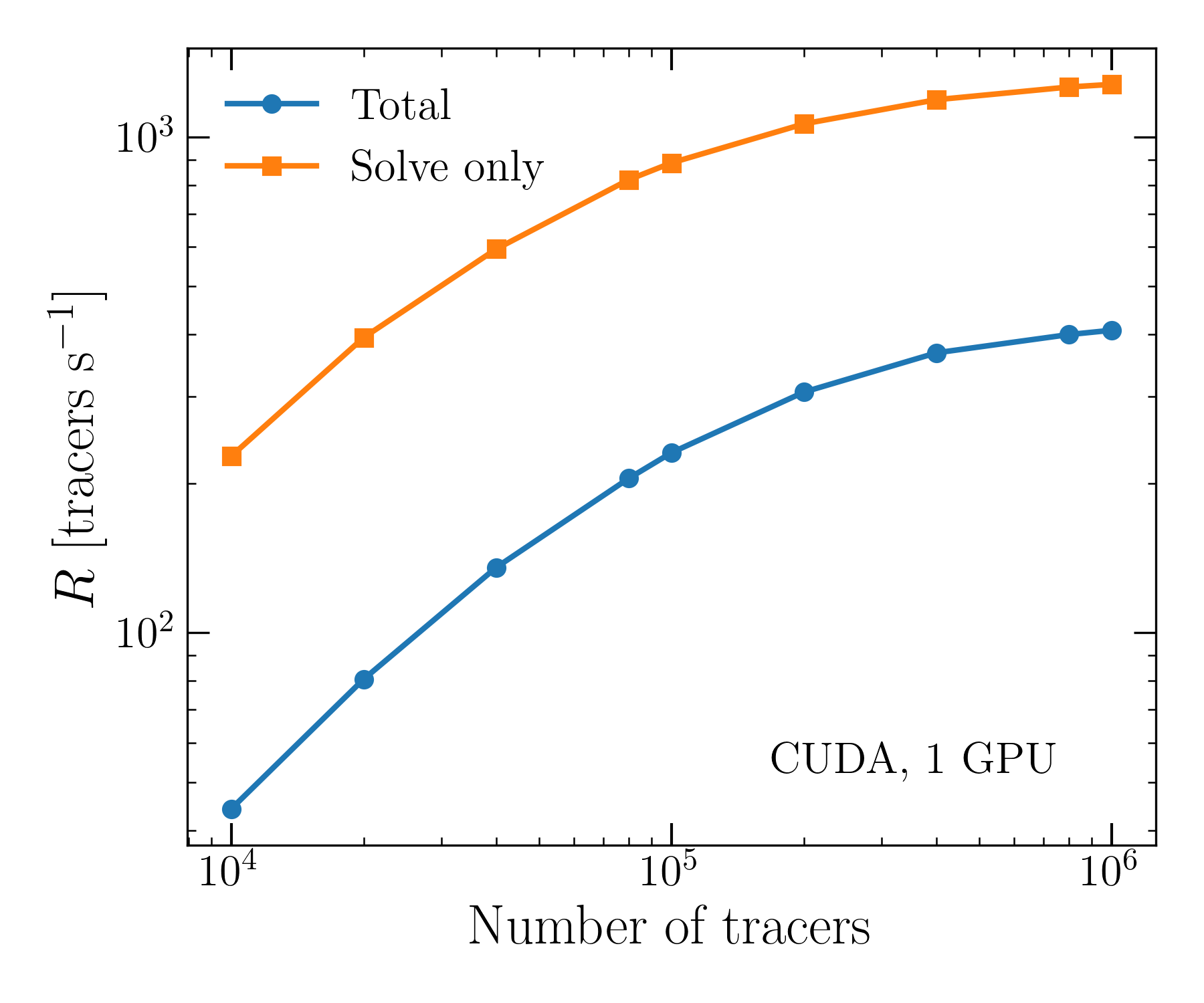}
    \caption{Single-GPU throughput as a function of $N_{\rm trc}$ per rank. The blue and orange lines show the throughput measured from the total wall-clock time and from the time spent in the tridiagonal solve, respectively.}
    \label{fig:gpu_throughput}
\end{figure}

\begin{figure*}
    \centering
    \begin{minipage}{0.48\linewidth}
        \includegraphics[width=\linewidth]{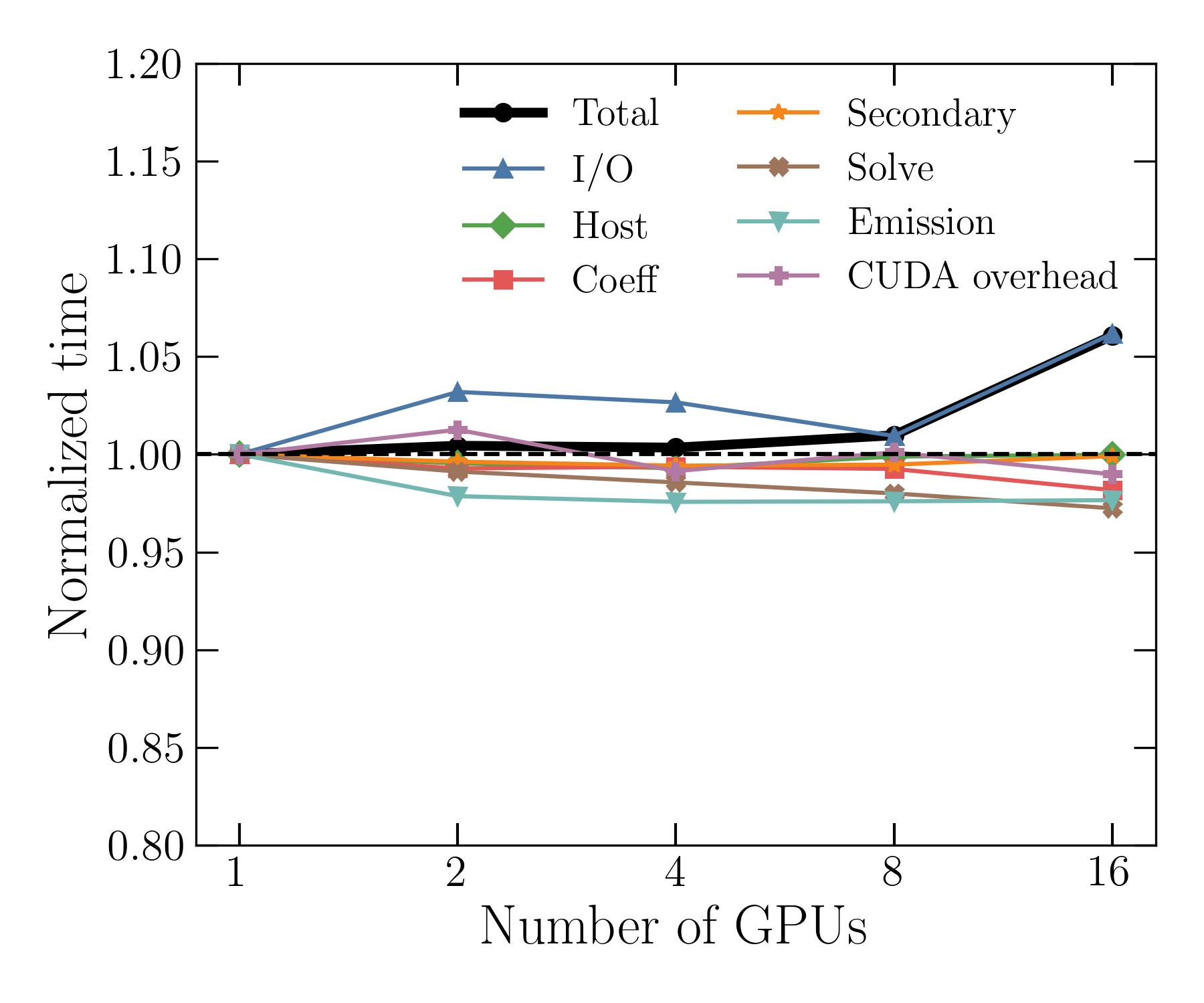}
    \end{minipage}
    \begin{minipage}{0.48\linewidth}
        \includegraphics[width=\linewidth]{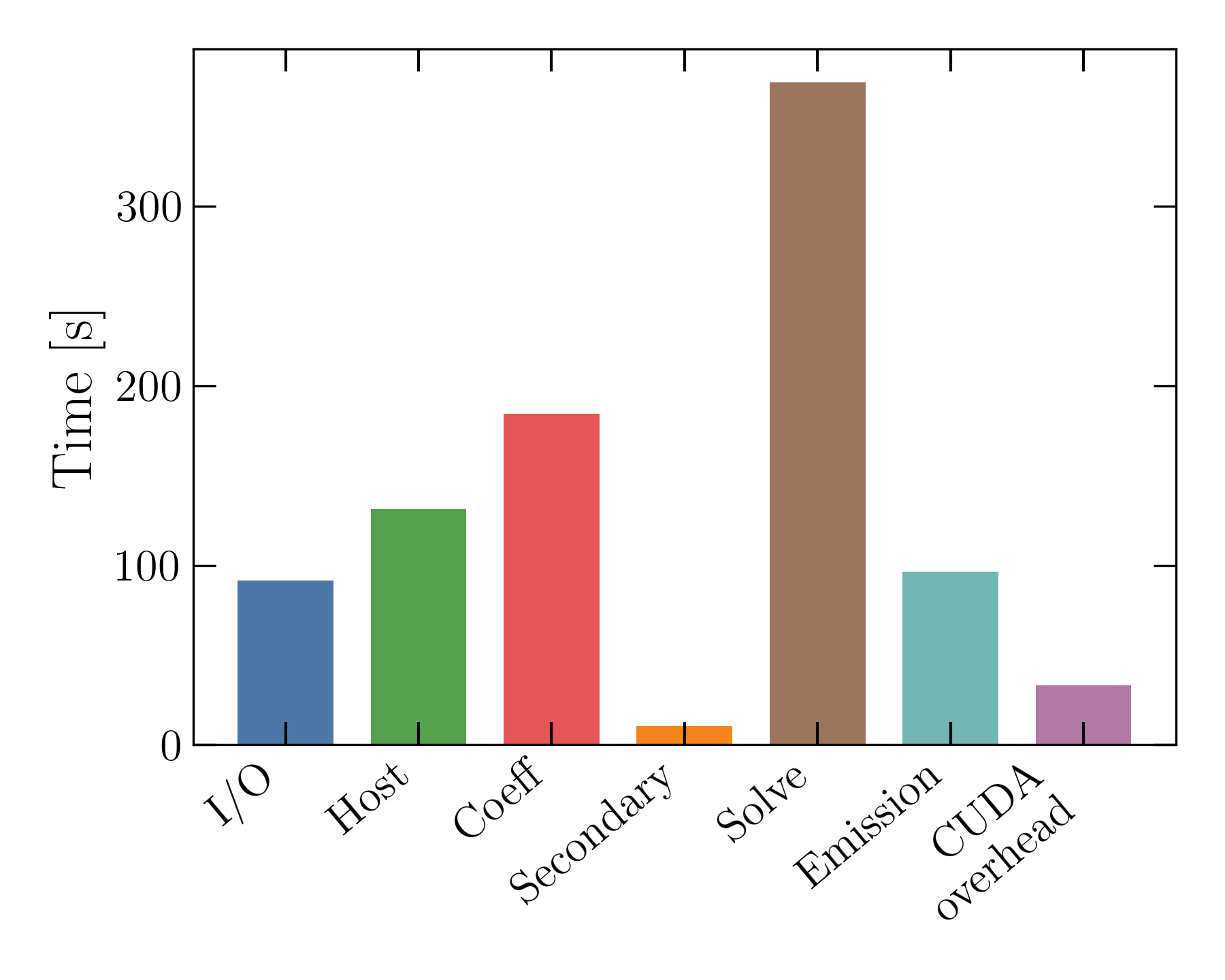}
    \end{minipage}
    \caption{GPU weak scaling test (left panel) and time breakdown (right panel). In the left panel, we show the total wall-clock time and component-wise execution times as a function of the number of GPUs. The thick black line shows the total wall time, while the thin lines show the time spent in each process, averaged over MPI ranks. 
    We use a fixed number of $5\times10^5$ tracers per rank. 
    In the right panel, we show the time breakdown for the 16-GPU run in the weak-scaling test.
    }
    \label{fig:gpu_weak}
\end{figure*}

We next evaluate the performance obtained with the CUDA/GPU backend using the PCR tridiagonal solver.
We measure the single-GPU throughput as a function of the problem size.
Specifically, we vary the number of tracers per rank, $N_{\rm trc}$, and examine the saturation of the throughput in single-MPI, one-GPU runs.
\par

Fig.~\ref{fig:gpu_throughput} shows the single GPU throughput for $10^4 \leq N_{\rm trc} \leq 2\times 10^6$. 
The blue line shows the end-to-end throughput, computed with the wall-clock time $T = T_{\rm wall}$, which includes I/O, preparation of FP coefficients, execution of the PCR solver, calculation of synchrotron emission, and other host- and device-side overheads.
The orange line shows the throughput of the tridiagonal solve alone, computed with $T = T_{\rm solve}$.
In our test case with $N_{\rm bin}=128$, the throughput of the tridiagonal solver saturates at around $R_{\rm solve}\approx10^3$ tracers s$^{-1}$ per MPI task.
We find that the efficiency of GPU acceleration increases with $N_{\rm trc}$, and the end-to-end throughput reaches $R_{\rm wall} \approx400$ tracers s$^{-1}$ per MPI task at $N_{\rm trc}\sim10^6$.
This is almost two orders of magnitude larger than the single-rank CPU throughput (Fig.~\ref{fig:cpu_strong}).
\par

Fig.~\ref{fig:gpu_throughput} shows that small-sized problems, such as those with $N_{\rm trc}\sim10^4$ tracers per task, cannot fully utilize the capability of the GPU.
A suitable use case for the CUDA/GPU backend is therefore $N_{\rm trc}\sim10^5 -10^6$ tracers per rank in the case of $N_{\rm bin}=128$.
However, one should also take care of the increase of the device memory consumption with $N_{\rm trc}$.
Including the arrays for the CR spectra and the FP coefficients, the total array size per tracer is approximately 32 KiB for $N_{\rm bin}=128$.
Using a single A100 GPU with 64 GB of memory on Leonardo, we could not run the case with $N_{\rm trc}=2\times10^6$ because of device memory overflow.
We note that the preferable range of $N_{\rm trc}$ per task depends on $N_{\rm bin}$.
As $N_{\rm bin}$ increases, the computational and memory costs per tracer increase, so a smaller number of tracers per task can be sufficient to make effective use of the GPU.
In practice, values of $N_{\rm bin}=128$, 256, or 512 are reasonable choices depending on the available computational resources and the required numerical accuracy (see also \ref{app:CC}).
\par


In the left panel of Fig.~\ref{fig:gpu_weak}, we assess the weak scaling of the GPU code. 
We test the cases of 1, 2, 4, 8 and 16 GPUs with $N_{\rm trc}=$ $5\times10^5$, $10^6$, $2\times10^6$, $4\times10^6$, and $8\times10^6$, respectively.
The number of tracers per rank is fixed at $5\times10^5$.
Since each computational node of the Leonardo Booster system is equipped with four GPUs and 32 CPU cores, the runs with 8 and 16 GPUs are executed on two and four nodes, respectively.
\par

We find that the normalized total wall-clock time, shown by the thick black line, varies by only several percent over the range from 1 to 16 GPUs.
The right panel of Fig.~\ref{fig:gpu_weak} shows the breakdown of the execution time for the 16-GPU run.
We place timers in several parts of the code and decompose the execution time into seven categories: input of the tracer data and output of the CR and emission spectra (I/O); host-side operations in the snapshot loop (Host); calculation of the FP and cooling coefficients (Coeff); calculation of the hadronic secondary kernels and secondary CR electron spectra (Secondary); tridiagonal solves (Solve); calculation of synchrotron and hadronic gamma-ray emissivities using the tabulated methods (Emission); and overhead associated with host--device communication and CUDA calls (CUDA overhead).
The breakdown is obtained by averaging the measured times over the 16 MPI ranks.
\par

We find that the off-loaded tridiagonal solve is the most time-consuming part (right panel of Fig.~\ref{fig:gpu_weak}).
The host-side operations, which mainly involve planning the tracer buckets and organizing the bucket arrays, together with the CUDA overhead, remain subdominant compared with the main computational costs associated with the coefficient calculation and tridiagonal solves.
As shown in the left panel, the Coeff and Solve components remain nearly constant with increasing GPU count, indicating that the main GPU-offloaded computations scale efficiently.
Overall, the GPU backend shows good weak scaling up to 16 GPUs, with only a modest increase in the total wall-clock time.

%

\section{Discussion and conclusions}
\label{sec:conclusions}

In this work, we present a C/C++-based parallel FP solver with a CUDA backend for GPU acceleration.
The code follows the spectral evolution of CRs and computes the associated non-thermal emission in post-processing using tracer-particle data from MHD simulations.
The physical model implemented in the code includes turbulent reacceleration and secondary-particle production through hadronic $pp$ collisions, which are plausible mechanisms for diffuse synchrotron emission in galaxy clusters.
\par

The code assumes Lagrangian tracer particles as input and neglects CR transport between tracers.
Under this assumption, the FP calculations for different tracers are independent, making the problem highly scalable.
The code exploits independence between tracers by performing batched FP solves for groups of tracers with the same number of FP sub-steps.
In the CUDA backend, the PCR solver parallelizes the tridiagonal solve over the momentum grid using CUDA threads, accelerating the computation compared with the classical serial Thomas solver adopted in the CPU backend.
The consistency between the CPU and GPU backends has been extensively tested in \ref{app:CC} and \ref{app:test}.
The potential bottleneck associated with the computation of synchrotron emissivity is avoided by using precomputed tables.
The accuracy of this tabulated method is presented in \ref{app:synch_table}.

We measure the code performance on a Leonardo Booster node equipped with 32 CPU cores and four NVIDIA A100 GPUs.
From the scaling test presented in Sect.~\ref{sec:cpu_strong}, we can see that the CPU backend reaches a per-node throughput of $\approx 120$ tracers s$^{-1}$ when using 32 MPI tasks for sufficiently large problem sizes.
On the other hand, the CUDA/GPU backend reaches $\approx 1600$ tracers s$^{-1}$
when using four GPUs (Figs.~\ref{fig:gpu_throughput} and \ref{fig:gpu_weak}).
Thus, the CUDA backend increases the effective per-node throughput by a factor of $\sim 10$.
\par

The speedup achieved with the CUDA implementation enables larger tracer calculations than would be practical with the CPU version alone. 
For example, in \citet{Nishiwaki26}, we studied emission in cluster outskirts using tracer particles with a characteristic mass of $\sim10^7 M_\odot$ and produced synthetic maps with an effective spatial resolution of $\approx16$ kpc.
The increase in per-node throughput by roughly a factor of ten obtained in this work would allow about ten times more tracers to be evolved at similar computational cost, enabling spatial sampling on scales of several kpc.
Such physical resolution would enable direct comparison of synthetic images with future SKA observations of nearby clusters at angular resolutions of a few arcseconds.

The speedup is also beneficial for exploring a substantially larger parameter space of uncertain sub-grid models, such as the re-acceleration efficiency and the CR proton-to-electron ratio. 
The code would help generate large suites of synthetic images and provide training or validation data for machine-learning approaches for radio observations \citep[e.g.,][]{Gheller_2018,Gheller_Vazza_2022,Sanvitale_2025}.
\par

The code can be extended in future work by including the effect of DSA in the CR source terms, and by coupling to cosmological MHD simulations through run-time FP solves.
The code could be ported to other GPU architectures, such as AMD GPUs, by implementing a portable GPU framework.


\section*{Acknowledgements}
We acknowledge INAF for awarding this project access to the LEONARDO supercomputer, owned by the EuroHPC Joint Undertaking, hosted by CINECA (Italy) and the LEONARDO consortium.
The computations were conducted using the resources allocated to the projects designated "INA24C6B13" and "INA24C7B01," with K.N. serving as the Principal Investigator.
We acknowledge financial support from INAF grants "Theory and simulations of non-thermal phenomena in galaxy clusters and beyond," and "A new numerical view of the cycle of relativistic electrons in clusters of galaxies."

\appendix

\section{The PCR algorithm}
\label{app:PCR}

\begin{lstlisting}[
    language=C++,
    caption={Schematic CUDA implementation of the PCR solver.
    One thread block is assigned to each tracer, and one thread
    is assigned to each momentum bin.},
    label={lst:pcr_cuda}
]
__global__ void solve_cc_pcr_kernel(...)
{
    int tracer = blockIdx.x;
    int j      = threadIdx.x;
    int idx    = tracer * Nbin + j;

    // Initialize PCR coefficients 
    a[j] = -a_batch[idx];
    b[j] =  b_batch[idx];
    c[j] = -c_batch[idx];
    d[j] = dt * source[idx] + spectrum[idx];

    __syncthreads();

    for (int stride = 1; stride < Nbin; stride *= 2) {

        a_new[j] = 0.0;
        b_new[j] = b[j];
        c_new[j] = 0.0;
        d_new[j] = d[j];

        if (j - stride >= 0) {
            alpha = -a[j] / b[j - stride];

            a_new[j]  = alpha * a[j - stride];
            b_new[j] += alpha * c[j - stride];
            d_new[j] += alpha * d[j - stride];
        }

        if (j + stride < Nbin) {
            beta = -c[j] / b[j + stride];

            b_new[j] += beta * a[j + stride];
            c_new[j]  = beta * c[j + stride];
            d_new[j] += beta * d[j + stride];
        }

        __syncthreads();
        swap_buffers();
    }

    spectrum[idx] = d[j] / b[j];
}
\end{lstlisting}

We show a schematic CUDA kernel of the PCR solver in Listing~\ref{lst:pcr_cuda}.
The PCR loop starts by eliminating $x_{j-1}$ and $x_{j+1}$
in the $j$-th equation of the tridiagonal system using the $(j-1)$-th and $(j+1)$-th equations (Eq.~\ref{eq:tridiag}). The system equations after $m$ loops can be written as
\begin{align}
a_{j-s_{m}}^{[m]} x_{j-2s_m}
+ b_{j-s_{m}}^{[m]} x_{j-s_{m}}
+ c_{j-s_{m}}^{[m]} x_{j}
&= d_{j-s_{m}}^{[m]} , \label{eq:CyRe1}\\
a_j^{[m]} x_{j-s_m}
+ b_j^{[m]} x_j
+ c_j^{[m]} x_{j+s_m}
&= d_j^{[m]} , \label{eq:CyRe2} \\
a_{j+s_{m}}^{[m]} x_{j}
+ b_{j+s_{m}}^{[m]} x_{j+s_m}
+ c_{j+s_{m}}^{[m]} x_{j+2s_m}
&= d_{j+s_{m}}^{[m]} ,
\label{eq:CyRe3}
\end{align}
where
\begin{equation}
s_m = 2^m
\end{equation}
is the stride of the momentum index. The superscript $[m]$ denotes the coefficients after $m$ loops. 
To reconcile the conventional notation for the Chang--Cooper scheme with the PCR notation used in this section, we define the initial PCR coefficients as $a_j^{[0]} = -a_j$, $b_j^{[0]} = b_j$, $c_j^{[0]} = -c_j$, and $d_j^{[0]} = \Delta t\, Q_j + u_j^i$.
\par

By eliminating $x_{j-s_m}$ and $x_{j+s_m}$ in Eqs.(\ref{eq:CyRe1}),(\ref{eq:CyRe2}),and (\ref{eq:CyRe3}) with
\begin{align}
\alpha_j^{[m]} &=
-\frac{a_j^{[m]}}{b_{j-s_m}^{[m]}}, \\
\beta_j^{[m]} &=
-\frac{c_j^{[m]}}{b_{j+s_m}^{[m]}} ,
\end{align}
the updated coefficients are given by
\begin{align}
a_j^{[m+1]} &= \alpha_j^{[m]} a_{j-s_m}^{[m]}, \\
b_j^{[m+1]} &=
b_j^{[m]}
+ \alpha_j^{[m]} c_{j-s_m}^{[m]}
+ \beta_j^{[m]} a_{j+s_m}^{[m]}, \\
c_j^{[m+1]} &= \beta_j^{[m]} c_{j+s_m}^{[m]}, \\
d_j^{[m+1]} &=
d_j^{[m]}
+ \alpha_j^{[m]} d_{j-s_m}^{[m]}
+ \beta_j^{[m]} d_{j+s_m}^{[m]} .
\end{align}
At the boundaries, the corresponding neighboring equation may not exist.
In that case, the missing contribution is set to zero. Specifically, we set
$\alpha_j^{[m]}=0$ if $(j-s_m)<0$, and $\beta_j^{[m]}=0$ if
$(j+s_m) \ge N_{\rm bin}$. 
The terms involving coefficients outside the range $0 \le j < N_{\rm bin}$ are omitted.
\par

After $M = \log_2 N_{\rm bin} $ loops, the stride becomes $s_M = N_{\rm bin}$, and the system is reduced to independent equations,
\begin{equation}
b_j^{[M]} x_j = d_j^{[M]} .
\end{equation}
The solution of the tridiagonal system is given by
\begin{equation}
x_j = \frac{d_j^{[M]}}{b_j^{[M]}} .
\end{equation}
\par
PCR has a higher work complexity than the classical Thomas algorithm,
$O(N_{\rm bin}\log_2 N_{\rm bin})$ instead of $O(N_{\rm bin})$, but its fine-grained parallelism can make it faster on GPUs when large batches of independent tridiagonal systems are solved concurrently \citep[e.g.,][]{Zhang_PCR}.
\par

In our CUDA implementation, each tracer is assigned to one CUDA thread block. Within each PCR solve, the $N_{\rm bin}$ momentum bins of the tridiagonal system are mapped one-to-one to the threads within the block.
Each thread updates the coefficients corresponding to one momentum bin, and block-wide synchronizations are used between successive PCR iterations as the stride is doubled.
The block-wide synchronization ensures that all updated PCR coefficients are available before the next iteration begins (Listing~\ref{lst:pcr_cuda}).
\par

Since one momentum bin is mapped to one CUDA thread, increasing $N_{\rm bin}$ directly increases the number of threads per block.
It increases the shared-memory workspace required to store the PCR coefficients during the reduction. We adopt $N_{\rm bin}=512$ as a practical upper limit in the present implementation.
For unsupported grid sizes, the CUDA backend falls back to a device implementation of the Thomas algorithm.
\par

For the time integration of a tracer, FP sub-steps can be executed within a single fused CUDA kernel.
The CR proton and electron systems are solved sequentially within the same thread block and reuse the same shared-memory workspace.
Keeping the spectra resident on the device and performing successive FP updates within the kernel reduces repeated kernel-launch and host--device communication overheads.
The updated spectra are transferred back to host memory only after the device-side updates required for the current tracer batch have been completed.


\section{Accuracy of the synchrotron table}
\label{app:synch_table}

\begin{figure*}
 \centering
    \begin{minipage}{0.48\linewidth}
        \includegraphics[width=\linewidth]{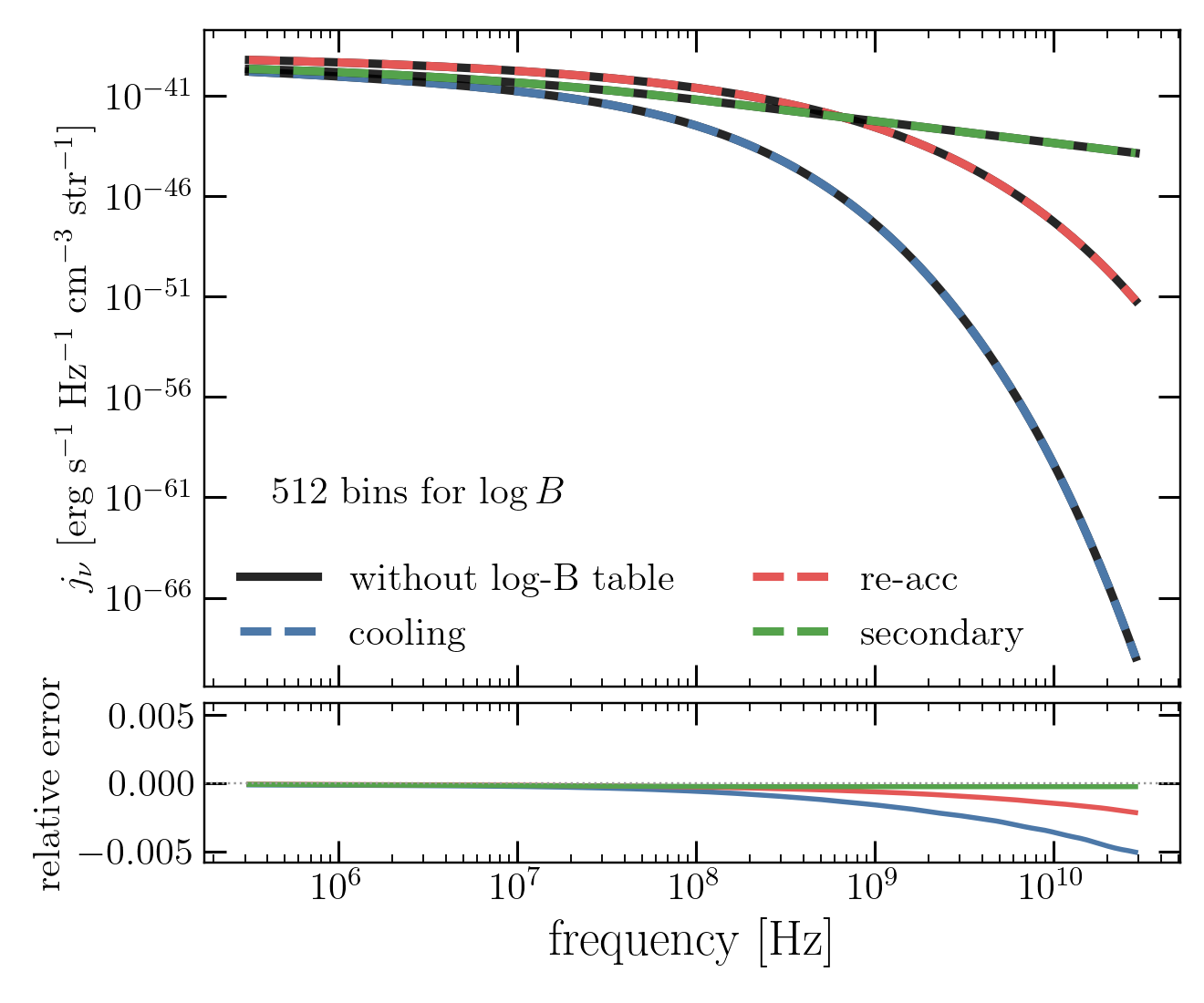}
    \end{minipage}
    \begin{minipage}{0.48\linewidth}
        \includegraphics[width=\linewidth]{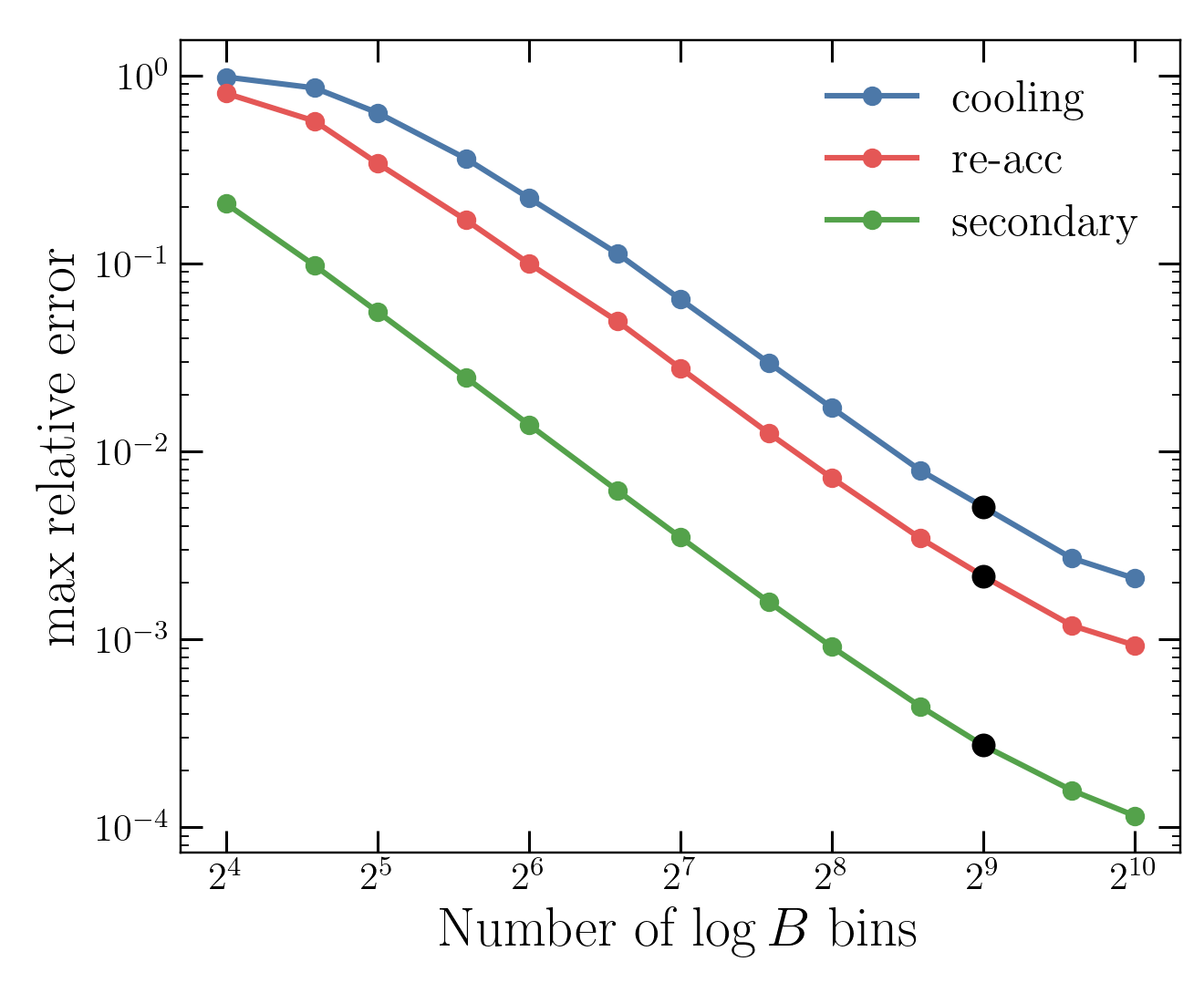}
    \end{minipage}
    \caption{Accuracy tests for synchrotron emission with the three models in Table \ref{tab:test}. In the left panel, we show the synchrotron spectra calculated with the tabulated method with the discretized magnetic field grid explained in Sect.~\ref{sec:synch} (colored lines) and compare them with the spectra calculated by direct numerical calculation of Eq.~(\ref{eq:synch}) without using the fixed-B tables (black lines). We use a fixed range for $-6 \leq \log_{10} B_{\mu G} \leq 2$ with 512 bins. The right panel shows the relation between the number of $\log_{10}B$ bins and the maximum relative error in the frequency range.}
    \label{fig:synch_table}
\end{figure*}

As explained in Sect.~\ref{sec:synch}, we use precomputated tables of Eq.~(\ref{eq:synch_table}) to speed up the calculation of the synchrotron integral (Eq.~\ref{eq:synch}).
In the left panel of Fig.~\ref{fig:synch_table}, we compare the spectra calculated with the $\log B$ table interpolation and those from direct numerical integration over the pitch-angle in Eq.~\ref{eq:synch}.
We use $N_\theta = 1024$ for the pitch-angle bin.
We use the default binning of the magnetic field with $-6 \leq \log_{10} B_{\mu G} \leq 2$ and 512 bins.
We study the cases of various spectral shapes, using the CR electron spectra calculated in the test of Sect.~\ref{app:test}.
We confirm that the error induced by the tabulated method is marginal except for the exponential tail at high frequencies.
In the right panel of Fig.~\ref{fig:synch_table}, we show the decay of the maximum relative error with the increased resolution of the $\log B$ bin.
The table interpolation method accelerates the calculation of the synchrotron emissivity by a factor of $N_\theta$, i.e., $\approx 10^3$.

\section{Comparison to analytical solutions}
\label{app:CC}
\begin{figure*}
    \centering
    \begin{minipage}{0.48\linewidth}
        \includegraphics[width=\linewidth]{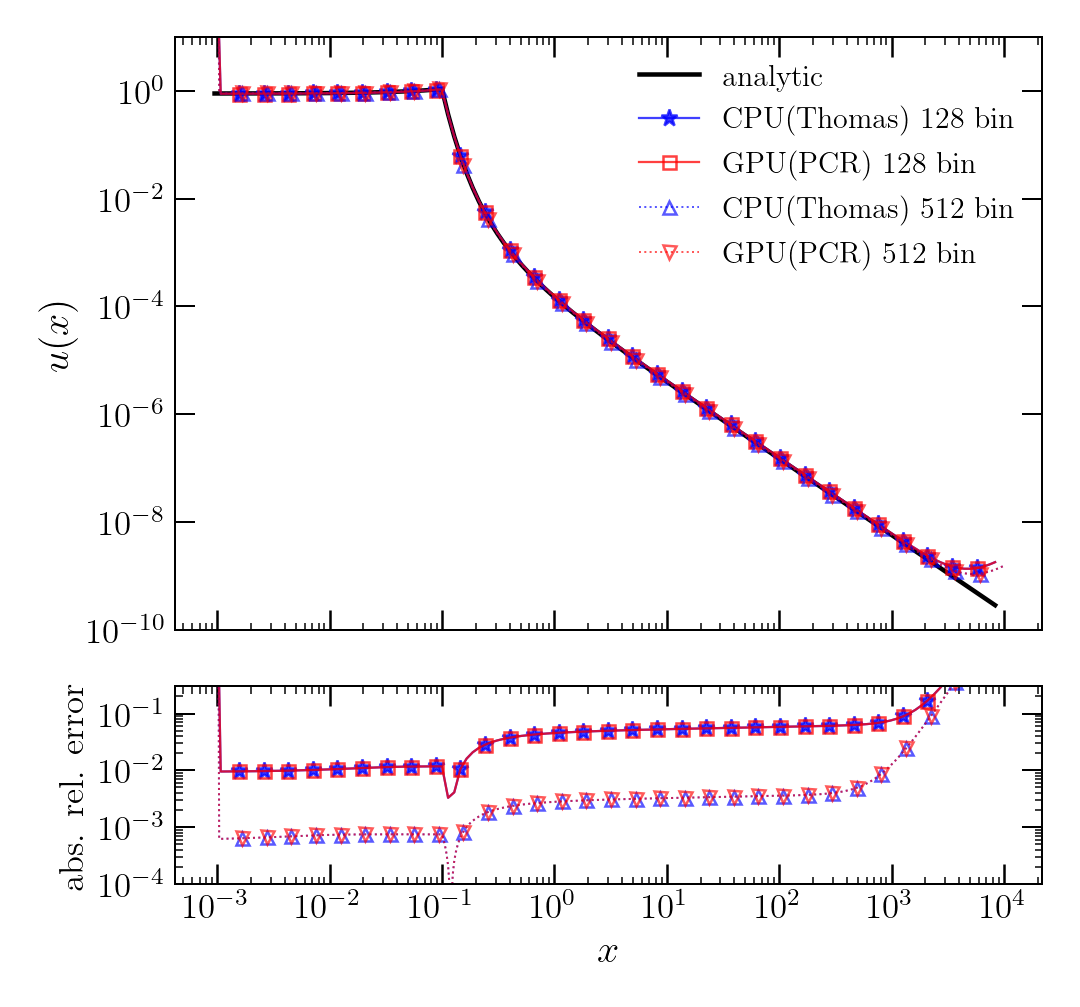}
    \end{minipage}
    \begin{minipage}{0.48\linewidth}
        \includegraphics[width=\linewidth]{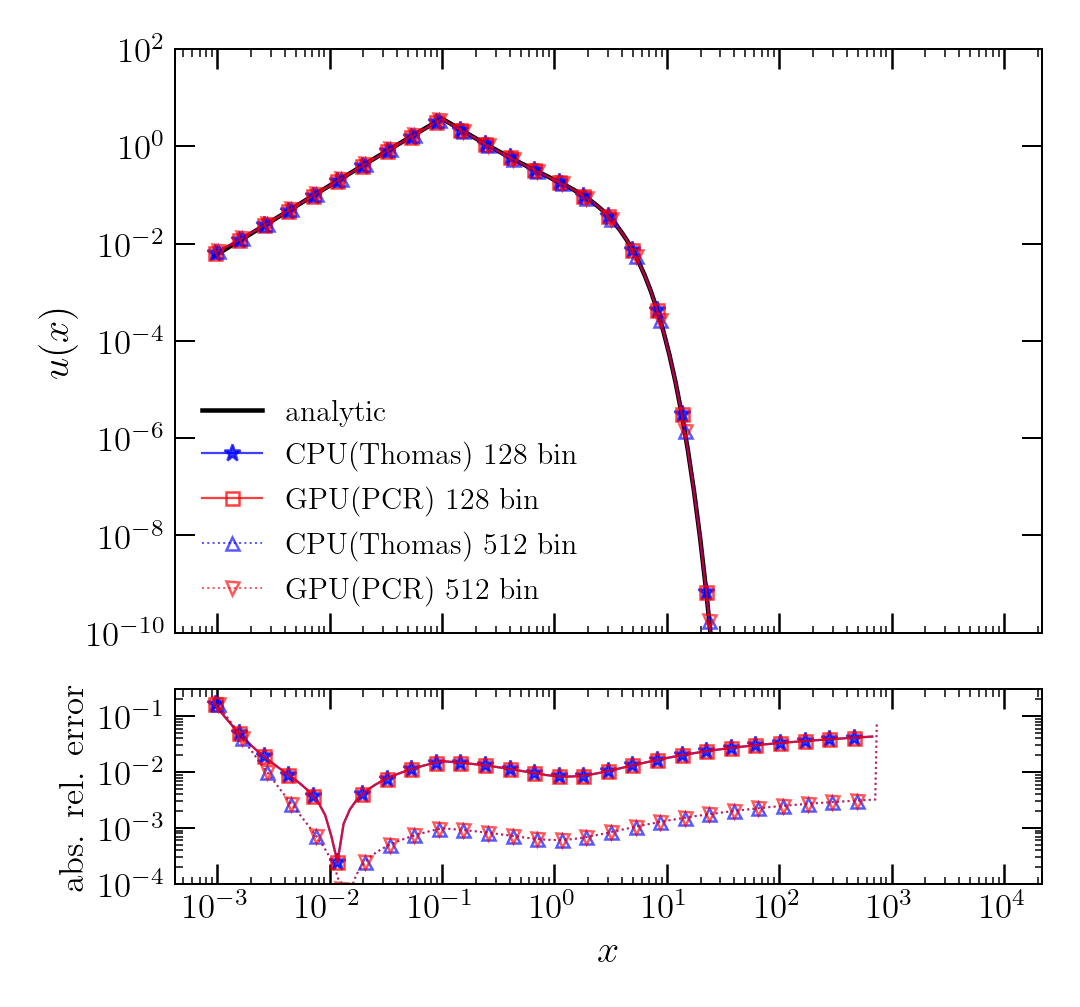}
    \end{minipage}
    \caption{Comparison to the analytical solutions of the hard-sphere problems. The upper panels show analytical (black) and numerical (blue and red) spectra. The left and right panels are the cases for $B(x) = -x+1$ and $B(x) = -x+x^2$, respectively. The solid lines with star or square markers show the results calculated with $N_{\rm bin} = 128$, while the dashed lines with triangle markers show $N_{\rm bin} = 512$ results. The lower panels show the absolute relative error to the analytical solutions.}
    \label{fig:analytic}
\end{figure*}

In this section, we test the implemented CC scheme and the tridiagonal solvers by comparing the numerical and analytical solutions of hard-sphere-type problems.
Here, we basically repeat the tests presented in \citet{Nishiwaki26} but with our newly developed CUDA pipeline.
\par

A generic form of FP equation can be written as
\begin{equation}
    \frac{\partial u}{\partial t} = \frac{1}{A(x)}\frac{\partial}{\partial x}\left[C(x)\frac{\partial u}{\partial x} + B(x)u\right] - \frac{u}{T(x)} + Q(x),
    \label{eq:FP_gen}
\end{equation}
where $x$ denotes the momentum or energy coordinate, $u(x,t)$ is the spectrum, $C(x)$, $B(x)$, $T(x)$, and $Q(x)$ are diffusive, advective, sink, and source coefficients, respectively.
We consider the hard-sphere-like case of $C(x) = x^2$, $T(x) = 1$, $A(x)=1$, and $Q(x) = \delta (x -x_{\rm inj})$ in Eq.~(\ref{eq:FP_gen}).
We consider two forms of the momentum-advection term: $B(x)=-x+1$, representing a case dominated by adiabatic expansion, and $B(x)=-x+x^2$, representing a case dominated by radiative cooling.
The minimum and maximum values of $x$ are set to $x_{\rm m} = 10^{-3}$ and $x_{\rm M} = 10^{4}$.
The steady state solutions for these problems can be found in \citet{Park_Petrosian_95}.
\par


Fig.~\ref{fig:analytic} shows the steady-state solutions for the above setup.
We compare two cases for the resolution of the $x$ bin, $N_x =$ 128 and 512.
The numerical solutions match well with the analytical ones in both cases, except for the bins close to the boundary.
The differences between CPU and GPU solutions are marginal compared with the discretization error inherent in the CC scheme.
These tests validate the tridiagonal solvers implemented in the code.



\section{Simulated spectra under ICM conditions}
\label{app:test}

\begin{figure*}
  \centering
  \begin{minipage}{0.31\textwidth}
    \centering
    \includegraphics[width=\linewidth]{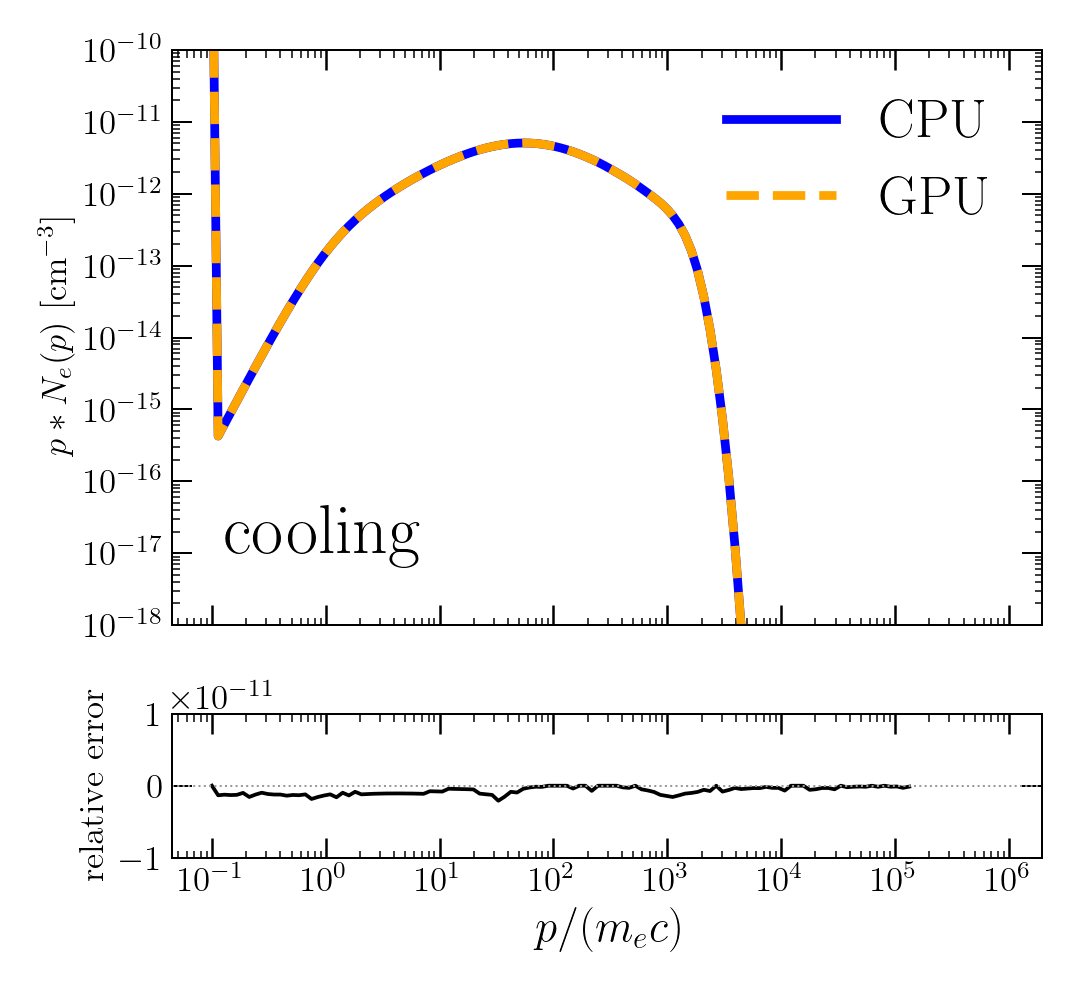}
  \end{minipage}
  \hfill
  \begin{minipage}{0.31\textwidth}
    \centering
    \includegraphics[width=\linewidth]{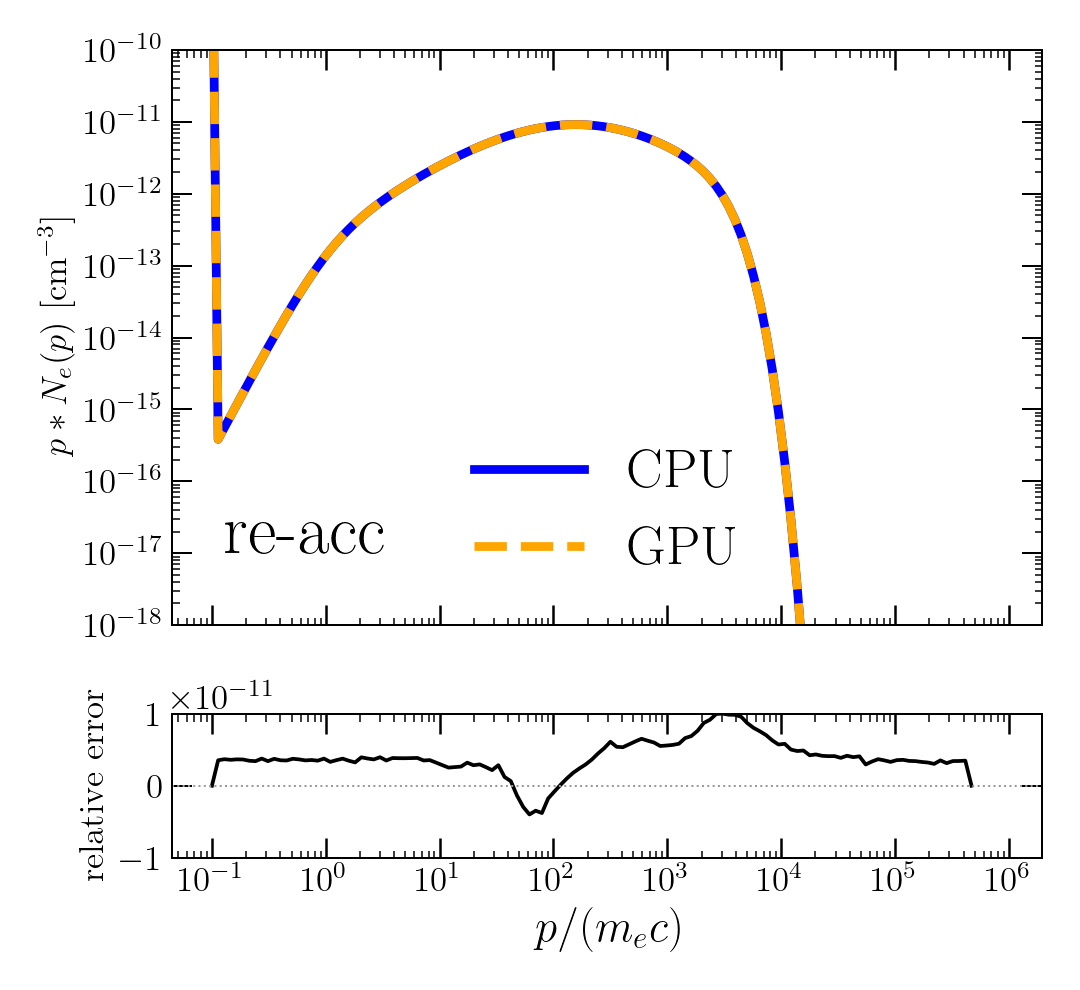}
  \end{minipage}
  \hfill
  \begin{minipage}{0.31\textwidth}
    \centering
    \includegraphics[width=\linewidth]{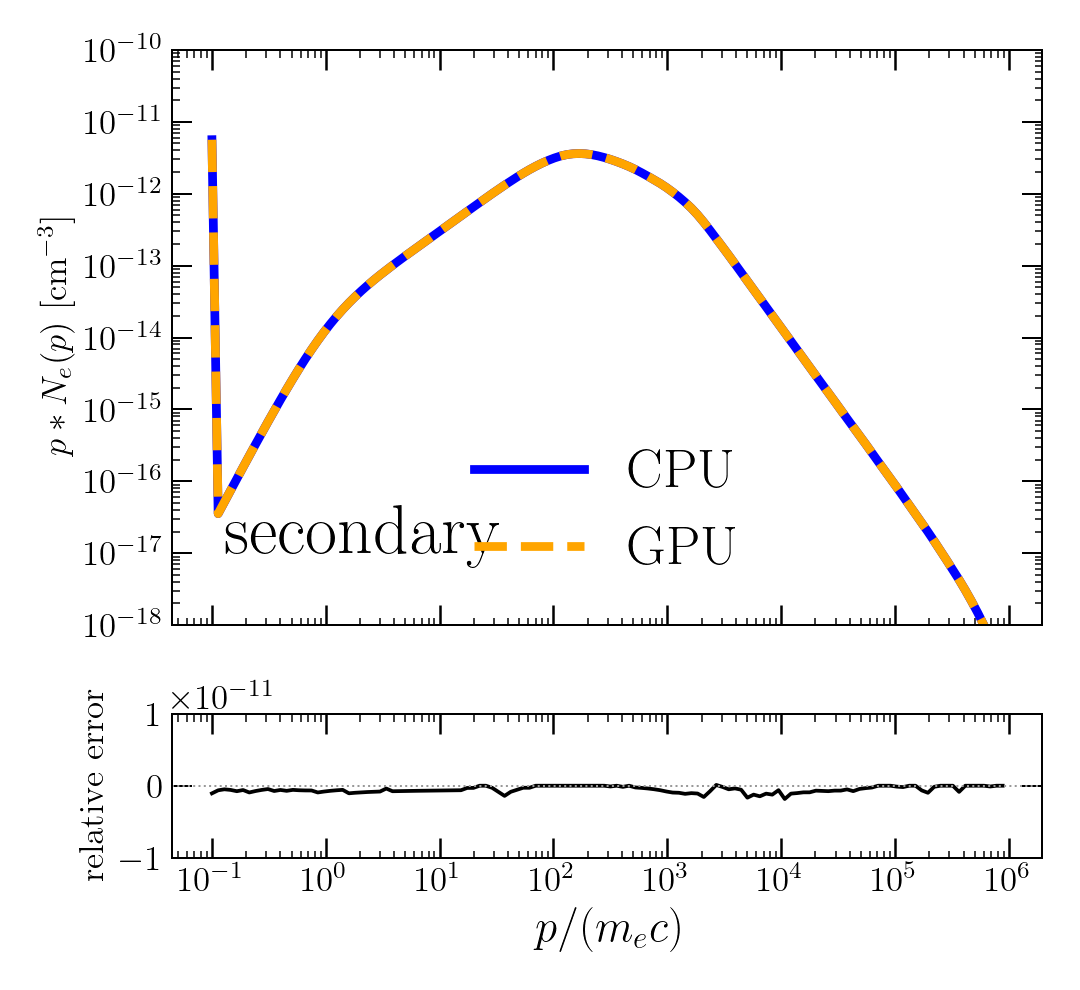}
  \end{minipage}
  \begin{minipage}{0.31\textwidth}
    \centering
    \includegraphics[width=\linewidth]{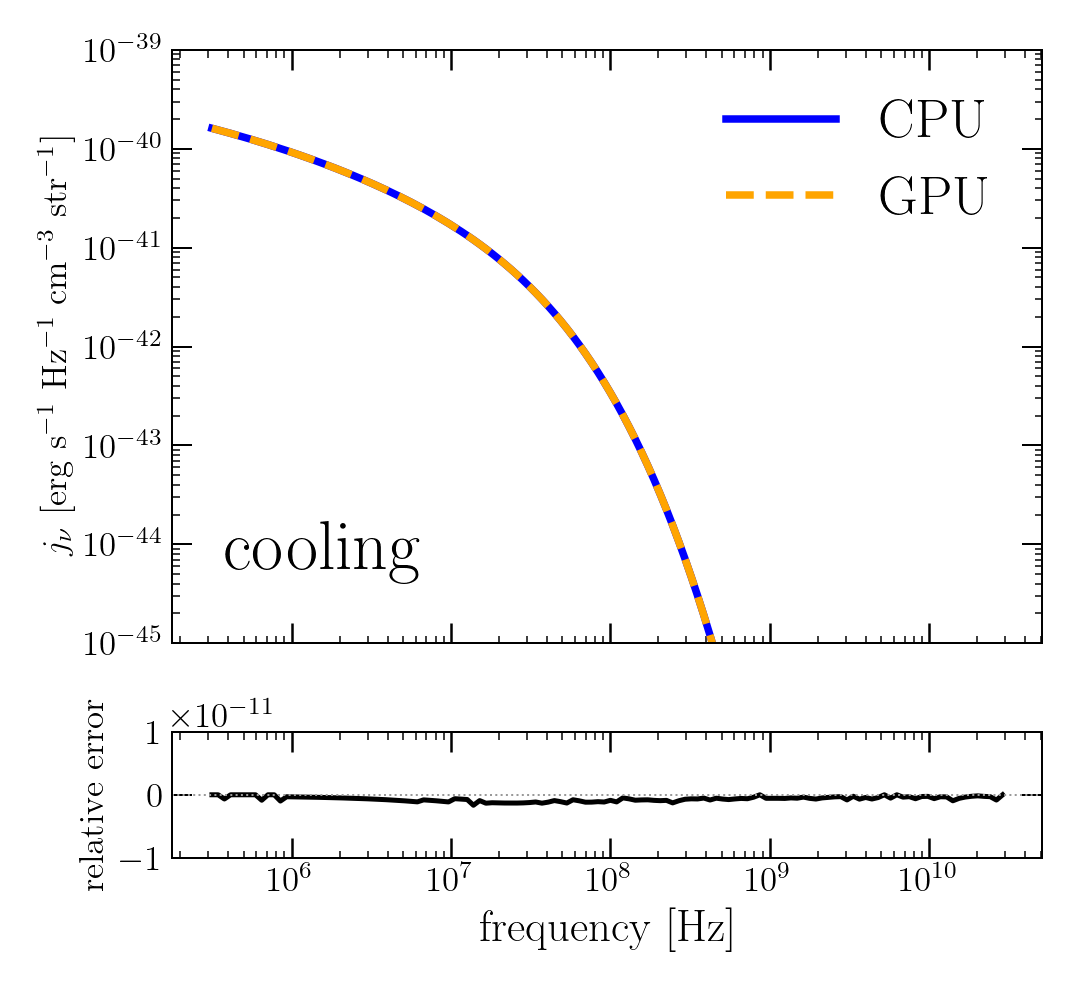}
  \end{minipage}
  \hfill
  \begin{minipage}{0.31\textwidth}
    \centering
    \includegraphics[width=\linewidth]{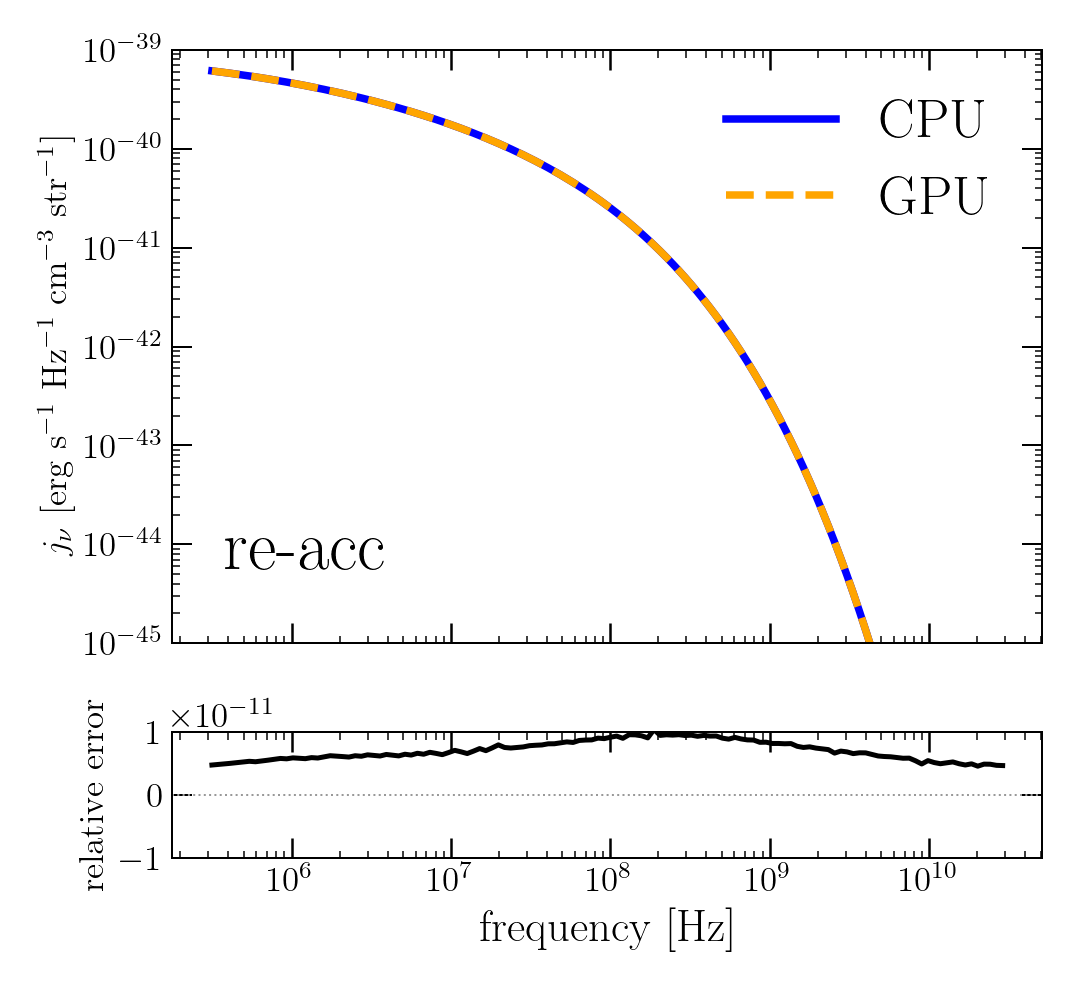}
  \end{minipage}
  \hfill
  \begin{minipage}{0.31\textwidth}
    \centering
    \includegraphics[width=\linewidth]{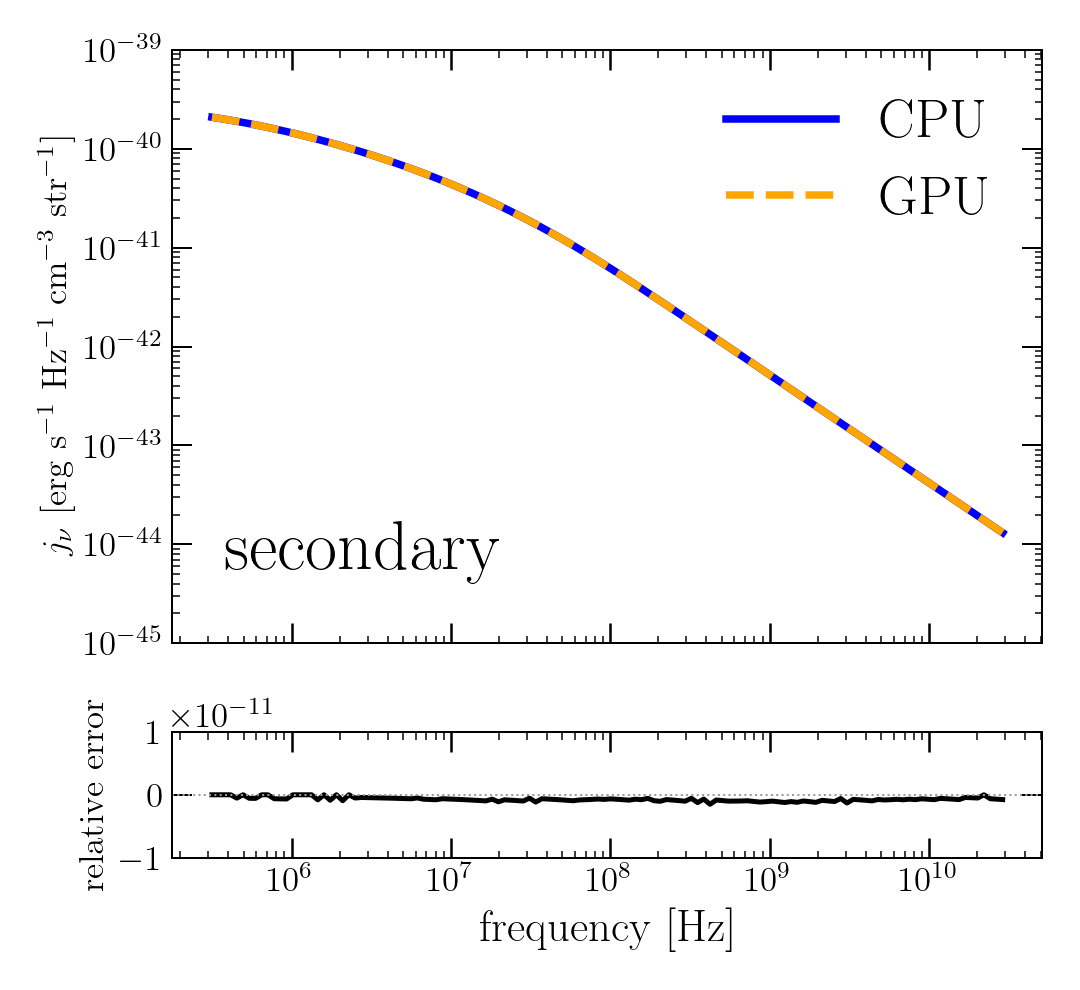}
  \end{minipage}
  \caption{
   CR electron (upper) and synchrotron (lower) spectra for the three models in Table \ref{tab:test}. 
  We also show the relative errors between solutions calculated with CPU and CUDA/GPU paths.
  }
  \label{fig:spectra}
\end{figure*}

\begin{table*}[]
    \centering
    \begin{tabular}{ccccccccc}
      model  & $N_{\rm bin}$ & redshift & $\rho_{\rm gas}$ [g cm$^{-3}$] & $B$ [$\mu$G] & $\delta_{\rm inj}$ & $\phi_{\rm CRe}$ & $\phi_{\rm CRp}$ & $t_{\rm reacc}$ [Myr]\\
      \hline
      cooling & 128 & 0.2 & $3\times10^{-27}$ & 3 & -2.2 & $10^{-6}$ & 0 & --\\
      re-acceleration & 128 & 0.2 &$3\times10^{-27}$ & 3 & -2.2 & $10^{-6}$ & 0 & 300 \\
      secondary  &  128 & 0.2 &$3\times10^{-27}$ & 3 & -2.2 & 0 & $10^{-7}$ & --\\
      \hline
    \end{tabular}
    \caption{Parameters used in the code tests in \ref{app:test}.}
    \label{tab:test}
\end{table*}

In \ref{app:CC}, we check the accuracy and consistency of the FP solver for both CPU and GPU backends using the hard-sphere type problems with known analytical solutions.
In this section, we examine the spectra calculated with CPU and GPU pipelines in more realistic problems to show the qualitative effects of the physical parameters and validate the consistency of the two backends.
We consider benchmark models with representative ICM parameters. 
We consider three representative cases: a cooling-dominated model, a re-acceleration model, and a secondary-electron model. 
We use the physical parameters and assumptions summarized in Table~\ref{tab:test}.
We assume a single power-law spectrum with a slope of $\delta_{\rm inj} = -2.2$ for the initial CR spectrum.
The initial spectrum is normalized with the CR number fraction $\phi$ to the number of thermal electrons or protons.
The inverse-Compton cooling is calculated with a fixed redshift $z = 0.2$, for simplicity.
Using the fixed background parameters, we solve the evolution for 3 Gyr.
We compute the CR spectra and the synchrotron emissivity, and compare the results from the CPU and CUDA backends.
\par

Fig.~\ref{fig:spectra} shows the CR and synchrotron spectra for the three cases.
In upper panels, we show the CR electron spectra multiplied by momentum, $p\,\mathcal{N}_e(p)$.
We reproduce the typical bump-shaped spectrum caused by Coulomb and radiative cooling in the cooling-dominated model.
In the re-acceleration model, the bump is broadened and the cut-off is shifted to higher energies, which leads to an increase in the synchrotron emissivity.
In the secondary-electron model without re-acceleration, the spectrum shows a single power-law shape above $\sim$GeV energies.
In each case, the relative errors in the CRe and synchrotron spectra between the CPU and GPU calculations are of order $\mathcal{O}(10^{-11})$.
These results confirm the consistency between the two backends.

\section*{Declaration of generative AI and AI-assisted technologies in the manuscript preparation process}

During the preparation of this work, the authors used ChatGPT GPT-5.5 and 5.6 to assist with improving the language, clarity, and readability of the manuscript. After using this tool, the authors reviewed and edited the content as needed and take full responsibility for the content of the publication.

\section*{Declaration of competing interest}

The authors declare that they have no known competing financial interests or personal relationships that could have appeared to influence the work reported in this paper.

\bibliographystyle{elsarticle-harv} 
\bibliography{ref}

@ARTICLE{Fujita_2003,
       author = {{Fujita}, Yutaka and {Takizawa}, Motokazu and {Sarazin}, Craig L.},
        title = "{Nonthermal Emissions from Particles Accelerated by Turbulence in Clusters of Galaxies}",
      journal = {\apj},
         year = 2003,
        month = feb,
       volume = {584},
       number = {1},
        pages = {190-202},
          doi = {10.1086/345599},
archivePrefix = {arXiv},
       eprint = {astro-ph/0210320},
 primaryClass = {astro-ph},
       adsurl = {https://ui.adsabs.harvard.edu/abs/2003ApJ...584..190F}
}

@article{BL07,
author = {Brunetti, G. and Lazarian, A.},
doi = {10.1111/j.1365-2966.2007.11771.x},
issn = {00358711},
journal = {Monthly Notices of the Royal Astronomical Society},
number = {1},
pages = {245--275},
title = {{Compressible turbulence in galaxy clusters: Physics and stochastic particle re-acceleration}},
volume = {378},
year = {2007}
}

@ARTICLE{BL16,
       author = {{Brunetti}, G. and {Lazarian}, A.},
        title = "{Stochastic reacceleration of relativistic electrons by turbulent reconnection: a mechanism for cluster-scale radio emission?}",
      journal = {\mnras},
         year = 2016,
        month = may,
       volume = {458},
       number = {3},
        pages = {2584-2595},
          doi = {10.1093/mnras/stw496},
archivePrefix = {arXiv},
       eprint = {1603.00458},
 primaryClass = {astro-ph.HE},
       adsurl = {https://ui.adsabs.harvard.edu/abs/2016MNRAS.458.2584B}
}

@article{Brunetti_Jones_review,
author = {Brunetti, Gianfranco and Jones, Thomas W.},
doi = {10.1142/S0218271814300079},
issn = {02182718},
journal = {International Journal of Modern Physics D},
number = {4},
pages = {1--76},
title = {{Cosmic rays in galaxy clusters and their nonthermal emission}},
volume = {23},
year = {2014}
}

@ARTICLE{Bryan_2014,
       author = {{Bryan}, Greg L. and {Norman}, Michael L. and {O'Shea}, Brian W. and {Abel}, Tom and {Wise}, John H. and {Turk}, Matthew J. and {Reynolds}, Daniel R. and {Collins}, David C. and {Wang}, Peng and {Skillman}, Samuel W. and {Smith}, Britton and {Harkness}, Robert P. and {Bordner}, James and {Kim}, Ji-hoon and {Kuhlen}, Michael and {Xu}, Hao and {Goldbaum}, Nathan and {Hummels}, Cameron and {Kritsuk}, Alexei G. and {Tasker}, Elizabeth and {Skory}, Stephen and {Simpson}, Christine M. and {Hahn}, Oliver and {Oishi}, Jeffrey S. and {So}, Geoffrey C. and {Zhao}, Fen and {Cen}, Renyue and {Li}, Yuan and {Enzo Collaboration}},
        title = "{ENZO: An Adaptive Mesh Refinement Code for Astrophysics}",
      journal = {\apjs},
         year = 2014,
        month = apr,
       volume = {211},
       number = {2},
          eid = {19},
        pages = {19},
          doi = {10.1088/0067-0049/211/2/19},
archivePrefix = {arXiv},
       eprint = {1307.2265},
 primaryClass = {astro-ph.IM},
       adsurl = {https://ui.adsabs.harvard.edu/abs/2014ApJS..211...19B}
}

@ARTICLE{Chang_Cooper_70,
       author = {{Chang}, J.~S. and {Cooper}, G.},
        title = "{A Practical Difference Scheme for Fokker-Planck Equations}",
      journal = {Journal of Computational Physics},
         year = 1970,
        month = aug,
       volume = {6},
       number = {1},
        pages = {1-16},
          doi = {10.1016/0021-9991(70)90001-X},
       adsurl = {https://ui.adsabs.harvard.edu/abs/1970JCoPh...6....1C}
}

@ARTICLE{Brunetti_2016,
       author = {{Brunetti}, Gianfranco},
        title = "{The challenge of turbulent acceleration of relativistic particles in the intra-cluster medium}",
      journal = {Plasma Physics and Controlled Fusion},
         year = 2016,
        month = jan,
       volume = {58},
       number = {1},
          eid = {014011},
        pages = {014011},
          doi = {10.1088/0741-3335/58/1/014011},
archivePrefix = {arXiv},
       eprint = {1509.03299},
 primaryClass = {astro-ph.HE},
       adsurl = {https://ui.adsabs.harvard.edu/abs/2016PPCF...58a4011B}
}

@ARTICLE{Park_Petrosian_96,
       author = {{Park}, Brian T. and {Petrosian}, Vahe},
        title = "{Fokker-Planck Equations of Stochastic Acceleration: A Study of Numerical Methods}",
      journal = {\apjs},
         year = 1996,
        month = mar,
       volume = {103},
        pages = {255},
          doi = {10.1086/192278},
       adsurl = {https://ui.adsabs.harvard.edu/abs/1996ApJS..103..255P}
}

@ARTICLE{vanWeeren_review,
       author = {{van Weeren}, R.~J. and {de Gasperin}, F. and {Akamatsu}, H. and {Br{\"u}ggen}, M. and {Feretti}, L. and {Kang}, H. and {Stroe}, A. and {Zandanel}, F.},
        title = "{Diffuse Radio Emission from Galaxy Clusters}",
      journal = {\ssr},
         year = 2019,
        month = feb,
       volume = {215},
       number = {1},
          eid = {16},
        pages = {16},
          doi = {10.1007/s11214-019-0584-z},
archivePrefix = {arXiv},
       eprint = {1901.04496},
 primaryClass = {astro-ph.HE},
       adsurl = {https://ui.adsabs.harvard.edu/abs/2019ssRv..215...16V}
}

@ARTICLE{Botteon_2022_A2255,
       author = {{Botteon}, Andrea and {van Weeren}, Reinout J. and {Brunetti}, Gianfranco and {Vazza}, Franco and {Shimwell}, Timothy W. and {Br{\"u}ggen}, Marcus and {R{\"o}ttgering}, Huub J.~A. and {de Gasperin}, Francesco and {Akamatsu}, Hiroki and {Bonafede}, Annalisa and {Cassano}, Rossella and {Cuciti}, Virginia and {Dallacasa}, Daniele and {Di Gennaro}, Gabriella and {Gastaldello}, Fabio},
        title = "{Magnetic fields and relativistic electrons fill entire galaxy cluster}",
      journal = {Science Advances},
         year = 2022,
        month = nov,
       volume = {8},
       number = {44},
          eid = {eabq7623},
        pages = {eabq7623},
          doi = {10.1126/sciadv.abq7623},
archivePrefix = {arXiv},
       eprint = {2211.01493},
 primaryClass = {astro-ph.CO},
       adsurl = {https://ui.adsabs.harvard.edu/abs/2022SciA....8.7623B}
}

@ARTICLE{Vazza_2023,
       author = {{Vazza}, F. and {Wittor}, D. and {Di Federico}, L. and {Br{\"u}ggen}, M. and {Brienza}, M. and {Brunetti}, G. and {Brighenti}, F. and {Pasini}, T.},
        title = "{Life cycle of cosmic-ray electrons in the intracluster medium}",
      journal = {\aap},
         year = 2023,
        month = jan,
       volume = {669},
          eid = {A50},
        pages = {A50},
          doi = {10.1051/0004-6361/202243753},
archivePrefix = {arXiv},
       eprint = {2210.01591},
 primaryClass = {astro-ph.HE},
       adsurl = {https://ui.adsabs.harvard.edu/abs/2023A&A...669A..50V}
}

@ARTICLE{Boess_2023,
       author = {{B{\"o}ss}, Ludwig M. and {Steinwandel}, Ulrich P. and {Dolag}, Klaus and {Lesch}, Harald},
        title = "{CRESCENDO: an on-the-fly Fokker-Planck solver for spectral cosmic rays in cosmological simulations}",
      journal = {\mnras},
         year = 2023,
        month = feb,
       volume = {519},
       number = {1},
        pages = {548-572},
          doi = {10.1093/mnras/stac3584},
archivePrefix = {arXiv},
       eprint = {2207.05087},
 primaryClass = {astro-ph.HE},
       adsurl = {https://ui.adsabs.harvard.edu/abs/2023MNRAS.519..548B}
}

@ARTICLE{Beduzzi_2024,
       author = {{Beduzzi}, L. and {Vazza}, F. and {Cuciti}, V. and {Brunetti}, G. and {Br{\"u}ggen}, M. and {Wittor}, D.},
        title = "{Cosmological simulations of the generation of cluster-scale radio emission from turbulent re-acceleration}",
      journal = {\aap},
         year = 2024,
        month = oct,
       volume = {690},
          eid = {A67},
        pages = {A67},
          doi = {10.1051/0004-6361/202450763},
archivePrefix = {arXiv},
       eprint = {2406.09859},
 primaryClass = {astro-ph.HE},
       adsurl = {https://ui.adsabs.harvard.edu/abs/2024A&A...690A..67B}
}

@ARTICLE{Rajpurohit_2021,
       author = {{Rajpurohit}, K. and {Brunetti}, G. and {Bonafede}, A. and {van Weeren}, R.~J. and {Botteon}, A. and {Vazza}, F. and {Hoeft}, M. and {Riseley}, C.~J. and {Bonnassieux}, E. and {Brienza}, M. and {Forman}, W.~R. and {R{\"o}ttgering}, H.~J.~A. and {Rajpurohit}, A.~S. and {Locatelli}, N. and {Shimwell}, T.~W. and {Cassano}, R. and {Di Gennaro}, G. and {Br{\"u}ggen}, M. and {Wittor}, D. and {Drabent}, A. and {Ignesti}, A.},
        title = "{Physical insights from the spectrum of the radio halo in MACS J0717.5+3745}",
      journal = {\aap},
         year = 2021,
        month = feb,
       volume = {646},
          eid = {A135},
        pages = {A135},
          doi = {10.1051/0004-6361/202039591},
archivePrefix = {arXiv},
       eprint = {2012.14373},
 primaryClass = {astro-ph.GA},
       adsurl = {https://ui.adsabs.harvard.edu/abs/2021A&A...646A.135R}
}

@ARTICLE{vanWeeren_2024,
       author = {{van Weeren}, R.~J. and {Timmerman}, R. and {Vaidya}, V. and {Gendron-Marsolais}, M. -L. and {Botteon}, A. and {Roberts}, I.~D. and {Hlavacek-Larrondo}, J. and {Bonafede}, A. and {Br{\"u}ggen}, M. and {Brunetti}, G. and {Cassano}, R. and {Cuciti}, V. and {Edge}, A.~C. and {Gastaldello}, F. and {Groeneveld}, C. and {Shimwell}, T.~W.},
        title = "{LOFAR high-band antenna observations of the Perseus cluster: The discovery of a giant radio halo}",
      journal = {\aap},
         year = 2024,
        month = dec,
       volume = {692},
          eid = {A12},
        pages = {A12},
          doi = {10.1051/0004-6361/202451618},
archivePrefix = {arXiv},
       eprint = {2410.02863},
 primaryClass = {astro-ph.CO},
       adsurl = {https://ui.adsabs.harvard.edu/abs/2024A&A...692A..12V}
}

@ARTICLE{Berezinsky_1997,
       author = {{Berezinsky}, V.~S. and {Blasi}, P. and {Ptuskin}, V.~S.},
        title = "{Clusters of Galaxies as Storage Room for Cosmic Rays}",
      journal = {\apj},
         year = 1997,
        month = oct,
       volume = {487},
       number = {2},
        pages = {529-535},
          doi = {10.1086/304622},
archivePrefix = {arXiv},
       eprint = {astro-ph/9609048},
 primaryClass = {astro-ph},
       adsurl = {https://ui.adsabs.harvard.edu/abs/1997ApJ...487..529B}
}

@ARTICLE{Voelk_1996,
       author = {{V{\"o}lk}, H.~J. and {Aharonian}, F.~A. and {Breitschwerdt}, D.},
        title = "{The Nonthermal Energy Content and Gamma-Ray Emission of Starburst Galaxies and Clusters of Galaxies}",
      journal = {\ssr},
         year = 1996,
        month = jan,
       volume = {75},
       number = {1-2},
        pages = {279-297},
          doi = {10.1007/BF00195040},
       adsurl = {https://ui.adsabs.harvard.edu/abs/1996SSRv...75..279V}
}

@ARTICLE{Peterson_2001,
       author = {{Peterson}, J.~R. and {Paerels}, F.~B.~S. and {Kaastra}, J.~S. and {Arnaud}, M. and {Reiprich}, T.~H. and {Fabian}, A.~C. and {Mushotzky}, R.~F. and {Jernigan}, J.~G. and {Sakelliou}, I.},
        title = "{X-ray imaging-spectroscopy of Abell 1835}",
      journal = {\aap},
         year = 2001,
        month = jan,
       volume = {365},
        pages = {L104-L109},
          doi = {10.1051/0004-6361:20000021},
archivePrefix = {arXiv},
       eprint = {astro-ph/0010658},
 primaryClass = {astro-ph},
       adsurl = {https://ui.adsabs.harvard.edu/abs/2001A&A...365L.104P}
}

@ARTICLE{Brunetti_2001,
       author = {{Brunetti}, G. and {Setti}, G. and {Feretti}, L. and {Giovannini}, G.},
        title = "{Particle reacceleration in the Coma cluster: radio properties and hard X-ray emission}",
      journal = {\mnras},
         year = 2001,
        month = jan,
       volume = {320},
       number = {3},
        pages = {365-378},
          doi = {10.1046/j.1365-8711.2001.03978.x},
archivePrefix = {arXiv},
       eprint = {astro-ph/0008518},
 primaryClass = {astro-ph},
       adsurl = {https://ui.adsabs.harvard.edu/abs/2001MNRAS.320..365B}
}

@ARTICLE{Nishiwaki_2021,
       author = {{Nishiwaki}, Kosuke and {Asano}, Katsuaki and {Murase}, Kohta},
        title = "{Particle Reacceleration by Turbulence and Radio Constraints on Multimessenger High-energy Emission from the Coma Cluster}",
      journal = {\apj},
         year = 2021,
        month = dec,
       volume = {922},
       number = {2},
          eid = {190},
        pages = {190},
          doi = {10.3847/1538-4357/ac1cdb},
archivePrefix = {arXiv},
       eprint = {2105.04541},
 primaryClass = {astro-ph.HE},
       adsurl = {https://ui.adsabs.harvard.edu/abs/2021ApJ...922..190N}
}

@ARTICLE{Park_Petrosian_95,
       author = {{Park}, Brian T. and {Petrosian}, Vahe},
        title = "{Fokker-Planck Equations of Stochastic Acceleration: Green's Functions and Boundary Conditions}",
      journal = {\apj},
         year = 1995,
        month = jun,
       volume = {446},
        pages = {699},
          doi = {10.1086/175828},
       adsurl = {https://ui.adsabs.harvard.edu/abs/1995ApJ...446..699P}
}

@ARTICLE{Dominguez-Fernandez_2026,
       author = {{Dom{\'\i}nguez-Fern{\'a}ndez}, P. and {ZuHone}, J.~A. and {Silich}, E.~M. and {Bellomi}, E. and {Sayers}, J. and {Mroczkowski}, T. and {Botteon}, A. and {van Weeren}, R.~J. and {Hernquist}, L. and {Brunetti}, G. and {Golec}, J. and {Gupta}, S.},
        title = "{ICM-SHOX III. The case of MACS J0018.5+1626, a radio relic that looks like a radio halo?}",
      journal = {arXiv e-prints},
         year = 2026,
        month = feb,
          eid = {arXiv:2602.05050},
        pages = {arXiv:2602.05050},
          doi = {10.48550/arXiv.2602.05050},
archivePrefix = {arXiv},
       eprint = {2602.05050},
 primaryClass = {astro-ph.HE},
       adsurl = {https://ui.adsabs.harvard.edu/abs/2026arXiv260205050D}
}

@ARTICLE{Ivleva_2026,
       author = {{Ivleva}, Anna and {B{\"o}ss}, Ludwig M. and {Dolag}, Klaus and {Koribalski}, B{\"a}rbel S. and {Khabibullin}, Ildar},
        title = "{Cosmological zoom-in simulation of odd radio circles as merger-driven shocks in galaxy groups}",
      journal = {\aap},
         year = 2026,
        month = feb,
       volume = {706},
          eid = {A80},
        pages = {A80},
          doi = {10.1051/0004-6361/202556716},
archivePrefix = {arXiv},
       eprint = {2508.00981},
 primaryClass = {astro-ph.GA},
       adsurl = {https://ui.adsabs.harvard.edu/abs/2026A&A...706A..80I}
}

@ARTICLE{Winner_2019,
       author = {{Winner}, Georg and {Pfrommer}, Christoph and {Girichidis}, Philipp and {Pakmor}, R{\"u}diger},
        title = "{Evolution of cosmic ray electron spectra in magnetohydrodynamical simulations}",
      journal = {\mnras},
         year = 2019,
        month = sep,
       volume = {488},
       number = {2},
        pages = {2235-2252},
          doi = {10.1093/mnras/stz1792},
archivePrefix = {arXiv},
       eprint = {1903.01467},
 primaryClass = {astro-ph.HE},
       adsurl = {https://ui.adsabs.harvard.edu/abs/2019MNRAS.488.2235W}
}

@ARTICLE{Nishiwaki26,
       author = {{Nishiwaki}, K. and {Brunetti}, G. and {Vazza}, F. and {Gheller}, C.},
        title = "{Cosmological simulation of a radio synchrotron bridge between pre-merging galaxy clusters}",
      journal = {\aap},
         year = 2026,
        month = may,
       volume = {709},
          eid = {A191},
        pages = {A191},
          doi = {10.1051/0004-6361/202558472},
archivePrefix = {arXiv},
       eprint = {2603.10187},
 primaryClass = {astro-ph.CO},
       adsurl = {https://ui.adsabs.harvard.edu/abs/2026A&A...709A.191N}
}

@inproceedings{Zhang_PCR,
author = {Zhang, Yao and Cohen, Jonathan and Owens, John D.},
title = {Fast tridiagonal solvers on the GPU},
year = {2010},
isbn = {9781605588773},
publisher = {Association for Computing Machinery},
address = {New York, NY, USA},
url = {https://doi.org/10.1145/1693453.1693472},
doi = {10.1145/1693453.1693472},
booktitle = {Proceedings of the 15th ACM SIGPLAN Symposium on Principles and Practice of Parallel Programming},
pages = {127–136},
numpages = {10},
location = {Bangalore, India},
series = {PPoPP '10}
}

@ARTICLE{Kamae_2006,
       author = {{Kamae}, Tuneyoshi and {Karlsson}, Niklas and {Mizuno}, Tsunefumi and {Abe}, Toshinori and {Koi}, Tatsumi},
        title = "{Parameterization of {\ensuremath{\gamma}}, e$^{+/-}$, and Neutrino Spectra Produced by p-p Interaction in Astronomical Environments}",
      journal = {\apj},
         year = 2006,
        month = aug,
       volume = {647},
       number = {1},
        pages = {692-708},
          doi = {10.1086/505189},
archivePrefix = {arXiv},
       eprint = {astro-ph/0605581},
 primaryClass = {astro-ph},
       adsurl = {https://ui.adsabs.harvard.edu/abs/2006ApJ...647..692K}
}

@ARTICLE{Kelner_2006,
       author = {{Kelner}, S.~R. and {Aharonian}, F.~A. and {Bugayov}, V.~V.},
        title = "{Energy spectra of gamma rays, electrons, and neutrinos produced at proton-proton interactions in the very high energy regime}",
      journal = {\prd},
         year = 2006,
        month = aug,
       volume = {74},
       number = {3},
          eid = {034018},
        pages = {034018},
          doi = {10.1103/PhysRevD.74.034018},
archivePrefix = {arXiv},
       eprint = {astro-ph/0606058},
 primaryClass = {astro-ph},
       adsurl = {https://ui.adsabs.harvard.edu/abs/2006PhRvD..74c4018K}
}

@ARTICLE{Kravtsov_Borgani_2012,
       author = {{Kravtsov}, Andrey V. and {Borgani}, Stefano},
        title = "{Formation of Galaxy Clusters}",
      journal = {\araa},
         year = 2012,
        month = sep,
       volume = {50},
        pages = {353-409},
          doi = {10.1146/annurev-astro-081811-125502},
archivePrefix = {arXiv},
       eprint = {1205.5556},
 primaryClass = {astro-ph.CO},
       adsurl = {https://ui.adsabs.harvard.edu/abs/2012ARA&A..50..353K}
}

@ARTICLE{Hussain_2023,
       author = {{Hussain}, Saqib and {Alves Batista}, Rafael and {de Gouveia Dal Pino}, Elisabete M. and {Dolag}, Klaus},
        title = "{The diffuse gamma-ray flux from clusters of galaxies}",
      journal = {Nature Communications},
         year = 2023,
        month = apr,
       volume = {14},
          eid = {2486},
        pages = {2486},
          doi = {10.1038/s41467-023-38226-w},
archivePrefix = {arXiv},
       eprint = {2203.01260},
 primaryClass = {astro-ph.HE},
       adsurl = {https://ui.adsabs.harvard.edu/abs/2023NatCo..14.2486H}
}

@ARTICLE{Fang_Murase_2018,
       author = {{Fang}, Ke and {Murase}, Kohta},
        title = "{Linking high-energy cosmic particles by black-hole jets embedded in large-scale structures}",
      journal = {Nature Physics},
         year = 2018,
        month = apr,
       volume = {14},
       number = {4},
        pages = {396-398},
          doi = {10.1038/s41567-017-0025-4},
archivePrefix = {arXiv},
       eprint = {1704.00015},
 primaryClass = {astro-ph.HE},
       adsurl = {https://ui.adsabs.harvard.edu/abs/2018NatPh..14..396F}
}

@ARTICLE{Gheller_2018,
       author = {{Gheller}, C. and {Vazza}, F. and {Bonafede}, A.},
        title = "{Deep learning based detection of cosmological diffuse radio sources}",
      journal = {\mnras},
         year = 2018,
        month = nov,
       volume = {480},
       number = {3},
        pages = {3749-3761},
          doi = {10.1093/mnras/sty2102},
archivePrefix = {arXiv},
       eprint = {1809.03315},
 primaryClass = {astro-ph.IM},
       adsurl = {https://ui.adsabs.harvard.edu/abs/2018MNRAS.480.3749G}
}

@ARTICLE{Gheller_Vazza_2022,
       author = {{Gheller}, C. and {Vazza}, F.},
        title = "{Convolutional deep denoising autoencoders for radio astronomical images}",
      journal = {\mnras},
         year = 2022,
        month = jan,
       volume = {509},
       number = {1},
        pages = {990-1009},
          doi = {10.1093/mnras/stab3044},
archivePrefix = {arXiv},
       eprint = {2110.08618},
 primaryClass = {astro-ph.IM},
       adsurl = {https://ui.adsabs.harvard.edu/abs/2022MNRAS.509..990G}
}

@ARTICLE{Sanvitale_2025,
       author = {{Sanvitale}, N. and {Gheller}, C. and {Vazza}, F. and {Bonafede}, A. and {Cuciti}, V. and {De Rubeis}, E. and {Govoni}, F. and {Murgia}, M. and {Vacca}, V.},
        title = "{Mapping diffuse radio sources using TUNA: a transformer-based deep learning approach}",
      journal = {\mnras},
         year = 2025,
        month = aug,
       volume = {541},
       number = {4},
        pages = {3479-3493},
          doi = {10.1093/mnras/staf1174},
archivePrefix = {arXiv},
       eprint = {2507.11320},
 primaryClass = {astro-ph.IM},
       adsurl = {https://ui.adsabs.harvard.edu/abs/2025MNRAS.541.3479S}
}

@ARTICLE{Blattnig_2000,
       author = {{Blattnig}, Steve R. and {Swaminathan}, Sudha R. and {Kruger}, Adam T. and {Ngom}, Moussa and {Norbury}, John W.},
        title = "{Parametrizations of inclusive cross sections for pion production in proton-proton collisions}",
      journal = {\prd},
         year = 2000,
        month = nov,
       volume = {62},
       number = {9},
          eid = {094030},
        pages = {094030},
          doi = {10.1103/PhysRevD.62.094030},
archivePrefix = {arXiv},
       eprint = {hep-ph/0010170},
 primaryClass = {hep-ph},
       adsurl = {https://ui.adsabs.harvard.edu/abs/2000PhRvD..62i4030B}
}

@book{gropp2014,
  author    = {Gropp, William and Lusk, Ewing and Skjellum, Anthony},
  title     = {Using MPI: Portable Parallel Programming with the Message-Passing Interface},
  edition   = {3rd},
  publisher = {MIT Press},
  address   = {Cambridge, MA},
  year      = {2014},
  isbn      = {978-0-262-52739-2}
}

@ARTICLE{Dolag_2008,
       author = {{Dolag}, K. and {Bykov}, A.~M. and {Diaferio}, A.},
        title = "{Non-Thermal Processes in Cosmological Simulations}",
      journal = {\ssr},
         year = 2008,
        month = feb,
       volume = {134},
       number = {1-4},
        pages = {311-335},
          doi = {10.1007/s11214-008-9319-2},
archivePrefix = {arXiv},
       eprint = {0801.1048},
 primaryClass = {astro-ph},
       adsurl = {https://ui.adsabs.harvard.edu/abs/2008SSRv..134..311D}
}

@ARTICLE{vanWeeren_2026_A1795,
       author = {{van Weeren}, R.~J. and {Osinga}, E. and {Brunetti}, G. and {Riseley}, C.~J. and {Botteon}, A. and {Timmerman}, R. and {Bonafede}, A. and {Br{\"u}ggen}, M. and {Cassano}, R. and {Cuciti}, V. and {Dallacasa}, D. and {de Gasperin}, F. and {de Jong}, J.~M.~G.~H.~J. and {Gastaldello}, F. and {Knowles}, K. and {Zhang}, X.},
        title = "{MeerKAT observations of Abell 1775 and Abell 1795: the discovery of a hadronic radio halo?}",
      journal = {\mnras},
         year = 2026,
        month = feb,
       volume = {546},
       number = {2},
          eid = {stag054},
        pages = {stag054},
          doi = {10.1093/mnras/stag054},
archivePrefix = {arXiv},
       eprint = {2601.16288},
 primaryClass = {astro-ph.CO},
       adsurl = {https://ui.adsabs.harvard.edu/abs/2026MNRAS.546ag054V}
}

@ARTICLE{2025ApJ...989..155K,
       author = {{Kawashima}, Tomohisa and {Asano}, Katsuaki},
        title = "{High-energy Neutrino Emission from a Radiatively Inefficient Accretion Flow with a Three-dimensional General Relativistic Magnetohydrodynamic Simulation}",
      journal = {\apj},
         year = 2025,
        month = aug,
       volume = {989},
       number = {2},
          eid = {155},
        pages = {155},
          doi = {10.3847/1538-4357/adee96},
archivePrefix = {arXiv},
       eprint = {2503.22891},
 primaryClass = {astro-ph.HE},
       adsurl = {https://ui.adsabs.harvard.edu/abs/2025ApJ...989..155K}
}






\end{document}